\documentclass[journal, onecolumn,12pt, draftclsnofoot]{IEEEtran}
\usepackage{mathrsfs}

\usepackage[noadjust]{cite}
\usepackage{graphicx,color,overpic,psfrag}
\usepackage{amsmath, amssymb}
\usepackage{latexsym}
\usepackage{bm}
\usepackage{amssymb}
\usepackage{cases}
\usepackage{array}
\usepackage{fancyhdr}
\usepackage{setspace}

\ifCLASSOPTIONcompsoc
\usepackage[caption=false,font=normalsize,labelfont=sf,textfont=sf]{subfig}
\else
\usepackage[caption=false,font=footnotesize]{subfig}
\fi

\usepackage{url}
\usepackage{algpseudocode}
\usepackage{algorithm}
\usepackage{blkarray}
\usepackage{booktabs}

\usepackage{multirow}
\usepackage{dsfont}
\usepackage{tabularx}
\usepackage[table]{xcolor}

\usepackage{amsfonts}

\usepackage{letltxmacro}

\graphicspath{{figure/}}

\newcommand{\ra}[1]{\renewcommand{\arraystretch}{#1}}

\newtheorem{theorem}{Theorem}

\newtheorem{remark}{Remark}

\newcommand{\thref}[1]{Theorem \ref{#1}}

\newcommand{\figref}[1]{Fig. \ref{#1}}
\newcommand{\tabref}[1]{Table \ref{#1}}

\newcommand{\appref}[1]{Appendix \ref{#1}}
\newcommand{\secref}[1]{Section \ref{#1}}

\newcommand{\tr}[1]{\mathrm{tr}\left(#1\right)}
\newcommand{\diag}[1]{\mathsf{diag}\left\{#1\right\}}

\newcommand{\logdet}[1]{\mathrm{log}\left(\mathrm{det}\left(#1\right)\right)}

\newcommand{\intd}{\mathrm{d}}
\newcommand{\intdx}[1]{\intd #1}

\newcommand{\argmax}[1]{\mathop{\arg\max}\limits_{#1}}

\newcommand{\maxis}[1]{\mathop{\mathrm{max}}\limits_{#1}}

\newcommand{\thetabs}[2]{{\dnnot{\theta}{bs}}}

\newcommand{\equaa}{\mathop{=}^{(\mathrm{a})}}

\newcommand{\ba}{\mathbf{a}}

\newcommand{\bn}{\mathbf{n}}

\newcommand{\bx}{\mathbf{x}}
\newcommand{\by}{\mathbf{y}}
\newcommand{\bz}{\mathbf{z}}

\newcommand{\bA}{\mathbf{A}}

\newcommand{\bE}{\mathbf{E}}

\newcommand{\bG}{\mathbf{G}}
\newcommand{\bH}{\mathbf{H}}
\newcommand{\bI}{\mathbf{I}}

\newcommand{\bK}{\mathbf{K}}

\newcommand{\bQ}{\mathbf{Q}}
\newcommand{\bR}{\mathbf{R}}

\newcommand{\bV}{\mathbf{V}}

\newcommand{\bX}{\mathbf{X}}
\newcommand{\bY}{\mathbf{Y}}

\newcommand{\bbC}{\mathbb{C}}
\newcommand{\bbR}{\mathbb{R}}

\newcommand{\hatbR}{\widehat{\bR}}

\newcommand{\bzero}{\mathbf{0}}

\newcommand{\bLambda}{{\boldsymbol\Lambda}}
\newcommand{\bOmega}{{\boldsymbol\Omega}}

\newcommand{\bXi}{{\boldsymbol\Xi}}
\newcommand{\bPi}{{\boldsymbol\Pi}}
\newcommand{\bPhi}{{\boldsymbol\Phi}}
\newcommand{\bPsi}{{\boldsymbol\Psi}}

\newcommand{\dnnot}[2]{#1_{\mathrm{#2}}}

\newcommand{\barjmath}{\bar{\jmath}}

\newcommand{\ntb}{\notag\\}

\allowdisplaybreaks

\begin{document}
	\title{Network Massive MIMO Transmission Over Millimeter-Wave and Terahertz Bands: Mobility Enhancement and Blockage Mitigation}

\author{
Li~You, Xu~Chen, Xiaohang~Song, Fan~Jiang, Wenjin~Wang, Xiqi~Gao,~\IEEEmembership{Fellow,~IEEE}, and~Gerhard~Fettweis,~\IEEEmembership{Fellow,~IEEE}%
\thanks{
\emph{Corresponding author: Xiqi Gao.}
}
\thanks{This work will be presented in part at the IEEE International Conference on Communications, Dublin, Ireland, Jun. 2020 \cite{Chen20mmWave}.
}
\thanks{
Li You, Xu Chen, Wenjin Wang, and Xiqi Gao are with the National Mobile Communications Research Laboratory, Southeast University, Nanjing 210096, China, and also with the Purple Mountain Laboratories, Nanjing 211100, China (e-mail: liyou@seu.edu.cn; chen\_xu@seu.edu.cn; wangwj@seu.edu.cn; xqgao@seu.edu.cn).}
\thanks{
Xiaohang Song and Gerhard Fettweis are with the Vodafone Chair Mobile Communications Systems, Technische Universit\"{a}t Dresden, 01062 Dresden, Germany (e-mail: xiaohang.song@tu-dresden.de; gerhard.fettweis@tu-dresden.de).}
\thanks{
Fan Jiang is with the Laboratory for Information and Decision Systems (LIDS), Massachusetts Institute of Technology (MIT), Cambridge, MA 02139 USA (e-mail: fjiang@mit.edu).
}
}
	
	\maketitle
	
	\begin{abstract}
		Mobility and blockage are two critical challenges in wireless transmission over millimeter-wave (mmWave) and Terahertz (THz) bands. In this paper, we investigate network massive multiple-input multiple-output (MIMO) transmission for mmWave/THz downlink in the presence of mobility and blockage.
		Considering the mmWave/THz propagation characteristics, we first propose to apply per-beam synchronization for network massive MIMO to mitigate the channel Doppler and delay dispersion effects. Accordingly, we establish a transmission model.
		We then investigate network massive MIMO downlink transmission strategies with only the  statistical channel state information (CSI) available at the base stations (BSs), formulating the strategy design problem as an optimization problem to maximize the network sum-rate.
	    We show that the beam domain is favorable to perform transmission, and demonstrate that BSs can work individually when sending signals to user terminals.
	    Based on these insights, the network massive MIMO precoding design is reduced to a network sum-rate maximization problem with respect to beam domain power allocation. By exploiting the sequential optimization method and random matrix theory, an iterative algorithm with guaranteed convergence performance is further proposed for beam domain power allocation.
	    Numerical results reveal that the proposed network massive MIMO transmission approach with the statistical CSI can effectively alleviate the blockage effects and provide mobility enhancement over mmWave and THz bands.
	\end{abstract}


	\section{Introduction}

	   The shortage of global wireless spectrum resources at sub-6 GHz has prompted the exploration of millimeter-wave (mmWave) and Terahertz (THz) bands, where the spectrum is orders of magnitude higher than that in current cellular allocations  \cite{Rappaport13Millimeter,Xiao17Millimeter,Rappaport19_100G}.
       Thanks to the deployment of large-scale antenna arrays at the base stations (BSs), massive multiple-input multiple-output (MIMO) could serve a plurality of user terminals (UTs) over the same time/frequency resources, remarkably increasing the system spectral efficiency \cite{Marzetta10Noncooperative}. From a practical point of view, combining mmWave/THz transmission with massive MIMO is a widespread concern \cite{Zhang2019Multiple,Swindlehurst14Millimeter,Huang17mmW,You17BDMA,Wang18mmW,Ai17mmW,Gao17Fast,Rappaport19_100G}. The short wavelength in mmWave/THz signals allows only a small physical space to encapsulate a large antenna array. Meanwhile, high beamforming gains of massive MIMO can solve the problem of high propagation path loss in mmWave/THz transmission.
	
	   Utilizing mmWave/THz frequencies for wireless communications has received significant attention over the past few years. Communicating over mmWave/THz bands is not merely a matter of adjusting the carrier frequency to the corresponding bands. The process is much more complicated. First, due to the short wavelength of mmWave/THz signals, the diffraction efforts are negligible. Also, it is important to consider that, as the number of scattering clusters is relatively small, the propagation path loss over mmWave/THz bands exhibits different properties. In addition, mmWave/THz signals are highly sensitive to blockage due to high penetration loss \cite{Petro17blockage,Bai15mmW}, which needs to be considered in transmission design.
	   Furthermore, in high mobility scenarios, the Doppler spread for signal transmission over the mmWave/THz bands could be much larger, becoming a system implementation bottleneck to realize the promising vision of mmWave/THz communications \cite{Andrews17Modeling}.

	   Existing works in mmWave/THz communications have attempted to address these issues.
	   Extensive measurement results for mmWave/THz channels have been reported in \cite{Azar13_28G,Samimi13_28G,Han15THz,Ruisi18modeling,Huang20channelmodeling,RuiSi20mmWv2v}.
	   Detailed modeling methods for blockage effects and spatial propagation characteristics of the mmWave channels have been studied in \cite{Akdeniz14Millimeter}.
	   Based on the random shape theory, a statistic model was developed in \cite{Bai15mmW} to quantify the blockage effects in mmWave propagation.	
	   To handle the mobility issue in MIMO transmissions,
	   beam-based Doppler frequency synchronization for narrowband MIMO transmissions was suggested in \cite{Va17beam}. For wideband massive MIMO wireless communications over mmWave/THz bands, a per-beam synchronization (PBS) approach was proposed in \cite{You17BDMA} motivated by the beam domain channel dispersion properties, where the synchronization is performed over each receive beam. In particular, the Doppler frequency spread with PBS can be approximately reduced by a factor of the number of UT antennas compared with the conventional synchronization approaches. In addition, assuming a fixed antenna array aperture, the number of antennas that can be accommodated scales linearly with the carrier frequency. Thus, the utilization of the PBS scheme can effectively alleviate the Doppler effect of the mmWave/THz transmission system \cite{You17BDMA}.

	   Distinct from the existing multi-cell massive MIMO transmission in \cite{Zhang15multicellMIMO,Sun17MulcellBeam}, we investigate network massive MIMO transmission in this paper, allowing the BS to transmit signals to UTs in other cells\footnote{Network massive MIMO systems are different from cell-free massive MIMO systems where the access points and UTs are all equipped with only one antenna \cite{Ngo17cellfree}.} and the resultant diversity against the channel blockage over the mmWave/THz bands can be efficiently exploited. This can significantly alleviate the blockage effects while improving the transmission performance over mmWave/THz bands. However, network transmission will also pose emerging challenges, including the considerable overhead of information exchange between BSs and the increased complexity of transmission strategy design. We target these challenges in designing network massive MIMO transmission strategies in this paper.
	
       The performance of network massive MIMO transmission is highly dependent on the accuracy of the obtained channel state information (CSI). Most previous works on multi-cell transmission designs \cite{Xiang13Coordinated,Tervo18Distributed,Mandar16mmW,He15Energy} require the complete knowledge of the instantaneous CSI at the transmitter (CSIT).
       However, the acquisition of the instantaneous CSIT is usually challenging in massive MIMO downlink (DL), especially over the mmWave/THz bands.
       For instance, utilizing the uplink (UL) and DL channel reciprocity, DL CSI can be obtained via UL channel estimation in time-division duplex systems. However, the obtained DL CSI can be inaccurate due to, e.g., the calibration error of the radio frequency chains \cite{Choi14Downlink}.
       Even worse, in frequency-division duplex systems without channel reciprocity, the CSI feedback overhead scales with the number of transmit antennas when orthogonal pilot sequences are adopted. The huge amount overhead is unacceptable in practical mmWave/THz network massive MIMO systems.
	   Moreover, when the UTs are in high mobility \cite{Ruisi16High-speed}, the channel fluctuation behaves much severer, and the acquired CSI quickly becomes outdated.
       Compared with the instantaneous CSI, the statistical CSI, e.g., the spatial correlation and channel mean, varies more slowly and can be easily and accurately obtained by the BSs through, e.g., long-term feedback or covariance extrapolation \cite{Wang12Statistical,Khalilsarai18fdd}. Therefore, the statistical CSI can be exploited for wireless transmission designs \cite{You17BDMA,Sun17MulcellBeam,You16Channel,You20LEO}, especially in the considered mmWave/THz network massive MIMO transmission system.
	  	
	   Motivated by the above considerations, we investigate the network massive MIMO transmission strategy design for mmWave/THz systems with the availability of the statistical CSIT at the BSs. The major contributions of this paper are summarized as follows
	\begin{itemize}
		\item We establish the network massive MIMO channel model over mmWave/THz bands, and propose a PBS scheme in both time and frequency dimensions to mitigate the channel Doppler and delay dispersion. The utilization of the PBS scheme makes the transmission design over mmWave/THz bands nearly the same as that over the regular bands; as a result, even in high mobility scenarios, the problem formulation and the corresponding analysis of the mmWave/THz transmission can be simplified.
        \item We formulate the problem to design the optimal DL transmission strategy as the maximization of the network sum-rate based on the PBS scheme. We then derive the necessary condition of the optimal transmit covariance matrices. The necessary condition reveals that with a sufficiently large number of transmit antennas, the beam domain is favorable to the DL transmission and the individual signals to the UTs from multiple BSs are possible.
		\item We explore a novel low-complexity algorithm to solve the above formulated problem. This is done by transforming the original precoder design problem into a beam domain power allocation problem. We show that the new problem is a concave-convex problem, and it can be efficiently solved by exploiting the concave-convex procedure (CCCP). We further derive a deterministic equivalent (DE) of the optimization objective, which can be calculated in low complexity.
		\item We demonstrate the efficiency of the proposed low-complexity algorithm through the exploitation of the Karush-Kuhn-Tucker (KKT) conditions and numerical simulations. Specifically, we show that the solution of each CCCP iteration has a structure similar to the classical water-filling solution. Based on that, we further develop an iterative water-filling power allocation algorithm with guaranteed convergence performance. Extensive numerical simulations are also conducted and the results strongly validate the significant performance improvement of the proposed network massive MIMO transmission approach over existing approaches in mmWave/THz bands, especially in highly mobile scenarios.

	\end{itemize}
	
	The rest of this paper is organized as follows.
    In \secref{sec:system_model}, we establish the mmWave/THz massive MIMO channel model with blockage effects and the transmission signal model after PBS.
    We then investigate the network transmission strategy to maximize the network DL transmission sum-rate with the statistical CSIT only in \secref{sec:transmission}. Extensive numerical results are presented in \secref{sec:simulation}. Finally, the conclusions are drawn in \secref{sec:conclusion}.
	
	We adopt the following notations throughout the paper.
	Upper-case and lower-case boldface letters indicate matrices and column vectors, respectively.
	We adopt $\barjmath=\sqrt{-1}$ to denote the imaginary unit.
	We use ${\mathbf{I}}_N$ to denote the $N \times N$ identity matrix.
	The superscripts ${( \cdot )^T}$, ${( \cdot )^{ - 1}}$, and ${( \cdot )^H}$ represent the matrix transpose, inverse, and conjugate-transpose operations, respectively. The ensemble expectation, matrix trace, and determinant operations are represented by $\mathbb{E}\left\lbrace\cdot\right\rbrace$, $\tr{\cdot}$, and $\mathrm{det} ( \cdot )$, respectively.
	$\mathcal{N}(a,\xi^2)$ represents the real-valued Gaussian distribution with mean $a$ and covariance $\xi^2$.
    We use ${[{\bf{A}}]_{m,n}}$, $\left[\bA\right]_{i,:}$, $\left[\bA\right]_{:,j}$ and $[\mathbf{a}]_m$ to represent the $(m,n)$-th element, the $i$-th row, the $j$-th column of matrix $\bA$, and the $m$-th element of column vector $\mathbf{a}$, respectively.
    The operator $\diag{\bf{x}}$ is to form a diagonal matrix with its $m$-th diagonal element being $[\mathbf{x}]_m$.
    $\lfloor x\rfloor$ is the floor function to obtain the largest integer not greater than $x$. The operator
    $\langle \cdot \rangle_M$ denotes the modulo-$M$ operation.
    $\delta\left(\cdot\right)$ denotes the Dirac's delta function.
	The inequality ${\bf{A}} \succeq \bzero$ means that ${\bf{A}}$ is a Hermitian positive semi-definite matrix.
    The notation $[x]^+$ denotes $\max(x, 0)$. The symbols $\odot$ and $\otimes$ are reserved for the Hadamard and Kronecker products, respectively.
	The notation $\triangleq$ is used for definitions.
	
	\section{System Model}\label{sec:system_model}
	\subsection{System Setup}
		
    We consider a network massive MIMO DL cellular system consisting of $U$ cells, where each cell $u\in\mathcal{U}\triangleq \left\lbrace 1,\cdots,U \right\rbrace$ contains one $M_u$-antennas BS and $K_u$ multi-antenna UTs.\footnote{ Note that
    $U$ is finite in practical systems, due to the implementation complexity considerations. In our work, we assume that the signals sent by the BSs outside the network contribute to the effective noise \cite{Huh12Network}.
    }
    Denote the BS in cell $u$ and the $k$-th UT in cell $u$ as BS-$u$ and UT-$\left(k,u\right)$, respectively. The set of all UTs is denoted by $\mathcal{K}\triangleq \left\lbrace\left(k,u\right)|k=1\dots,K_u;u=1,\dots,U\right\rbrace$. Note that there is a central processor in the considered network DL transmission, and BS-$u\in\mathcal{U}$ can send signals to all UTs-$\left(k,u\right)\in\mathcal{K}$.		
	
    Assume that BS-$u$ is equipped with a uniform planar array (UPA) consisting of $M_u = M_u^{\mathrm{h}}\times M_u^{\mathrm{v}}$ antennas, where half-wavelength spacing is used in both horizontal and vertical directions. The numbers of antennas in horizontal and vertical directions are $M_u^{\mathrm{h}}$ and $M_u^{\mathrm{v}}$, respectively. The number of total BS antennas in the system is $M_\mathrm{tot}= \sum_{u=1}^{U}M_u$.\footnote{
    The BSs in the network act as a single distributed multi-antenna transmitter with $M_\mathrm{tot}$ antennas, perfectly coordinated by a central controller, serving all UTs in the network \cite{Huh12Network}.}
    In addition, each UT-$\left(k,u\right)$ is equipped with a uniform linear array (ULA) consisting of $N_{k,u}$ antennas with half-wavelength antenna spacing.
    	
	Orthogonal frequency division multiplexing (OFDM) modulation is adopted with $\dnnot{N}{us}$ sub-carriers, appended with the cyclic prefix (CP) of length $\dnnot{N}{cp}$. The OFDM symbol duration and CP duration are $\dnnot{T}{us}=\dnnot{N}{us}\times \dnnot{T}{s}$ and $\dnnot{T}{cp}=\dnnot{N}{cp}\times \dnnot{T}{s}$, respectively, where $\dnnot{T}{s}$ is the system sampling interval.
	
	\subsection{Channel Model}
	
	A three-state statistical model is adopted to describe the mmWave/THz massive MIMO channel states, where each propagation link can be in one of three states: line-of-sight (LOS), non line-of-sight (NLOS), and outage states.\footnote{Compared with the channel model over the regular bands where each link is in either a LOS or NLOS state, the adopted three-state model incorporates an additional outage state which can describe the channel blockage effect over the mmWave and THz bands \cite{Akdeniz14Millimeter,Sara17network}.}
	The corresponding probability functions of the channels between BS-$v$ and UT-$\left(k,u\right)$ in each state can be written as 	
	\begin{subequations}\label{out_LOS_NLOS}
		\begin{align}
		p_{\mathrm{out}}\left(d_{k,u,v}\right) &= \mathrm{max}\left(0,1-\mathrm{e}^{-a_{\mathrm{out}}d_{k,u,v}+b_{\mathrm{out}}}\right),\label{P_out}\\
		p_{\mathrm{LOS}}\left(d_{k,u,v}\right) &= \left(1-p_{\mathrm{out}}\left(d_{k,u,v}\right)\right)\mathrm{e}^{-a_{\mathrm{los}}d_{k,u,v}},\\
		p_{\mathrm{NLOS}}\left(d_{k,u,v}\right) &= 1-p_{\mathrm{out}}\left(d_{k,u,v}\right)-p_{\mathrm{LOS}}\left(d_{k,u,v}\right),
		\end{align}
	\end{subequations}
	respectively, where $d_{k,u,v}$ is the distance between the BS and the UT, and $a_{\mathrm{out}}$, $a_{\mathrm{los}}$, and $b_{\mathrm{out}}$ are parameters that fit from the data \cite{Akdeniz14Millimeter}.
	According to the channel measurement campaigns \cite{Akdeniz14Millimeter,Mandar16mmW}, the large-scale path loss from BS-$v$ to UT-$\left(k,u\right)$ in a given state can be modeled via a standard linear form as
	\begin{align}\label{pathloss}
	\zeta_{k,u,v}^{\mathrm{sta}}\left(d_{k,u,v}\right)\left[\mathrm{in\ dB}\right] = \mathrm{a}^{\mathrm{sta}} + \mathrm{b}^{\mathrm{sta}} 10 \mathrm{log}_{10}\left(d_{k,u,v}\right)+ \mathrm{c}^{\mathrm{sta}},
	\end{align}
	where the superscript $\mathrm{sta}\in \left\lbrace \mathrm{LOS}, \mathrm{NLOS}, \mathrm{out} \right\rbrace$ represents the state of the link, $\mathrm{a}^{\mathrm{sta}}$, $\mathrm{b}^{\mathrm{sta}}$, and $\mathrm{c}^{\mathrm{sta}}$ are parameters that fit from the measurement results. In addition, $\mathrm{c}^{\mathrm{sta}}\in\mathcal{N}\left(0,(\xi^{\mathrm{sta}})^2\right)$ where $(\xi^{\mathrm{sta}})^2$ describes the shadowing variance.
	Note that these propagation state related parameters which represent large-scale fading vary relatively slowly compared to the instantaneous CSI. In the following, we model the channel parameters for a specific state where the propagation state index is omitted for notational brevity.
			
	According to the adopted configurations of the BS and UT antennas, steering vectors of UT-$\left(k,u\right)$ and BS-$v$ corresponding to the angles of arrival/departure (AoA/AoD) are given by
	\begin{subequations}
		\begin{align}
		\ba_{\mathrm{ut},k,u}\left(\theta\right)&=\left[1,\; \mathrm{e}^{-\barjmath\pi\theta},\;\ldots,\; \mathrm{e}^{-\barjmath\pi(N_{k,u}-1)\theta}\right]^{T}\in\bbC^{N_{k,u}\times 1},\\
		\ba_{\mathrm{bs},v}\left(\alpha,\beta\right)&=\left[1,\;\mathrm{e}^{-\barjmath\pi\alpha},\;\ldots, \mathrm{e}^{-\barjmath\pi(M_v^h-1)\alpha}\right]^{T}\otimes \left[1,\;\mathrm{e}^{-\barjmath\pi\beta},\;\ldots,\mathrm{e}^{-\barjmath\pi(M_v^v-1)\beta}\right]^{T} \in\bbC^{M_v\times 1},
		\end{align}
	\end{subequations}
	respectively, where $\theta$ is an auxiliary angle related to AoA $\vartheta\in\left[-\pi/2,\pi/2\right]$ as $\theta = \sin\vartheta\in\left[-1,1\right]$. $\alpha$ and $\beta$ are auxiliary angles related to elevation AoD $\phi_1\in\left[-\pi/2,\pi/2\right]$ and azimuth AoD $\phi_2\in\left[-\pi/2,\pi/2\right]$ as $\alpha = \cos\phi_2\sin\phi_1\in\left[-1,1\right]$ and $\beta=\sin\phi_2\sin\phi_1\in\left[-1,1\right]$, respectively.\footnote{Note that the array response vectors, $\ba_{\mathrm{ut},k,u}\left(\theta\right)$ and $\ba_{\mathrm{bs},v}\left(\alpha,\beta\right)$, can usually be assumed to be not related to the actual operating frequency in practical wideband mmWave/THz wireless systems as the bandwidth is usually much smaller compared with the carrier frequency \cite{Wang18mmW}.}
	Then, the baseband space domain channel response matrix from BS-$v$ to UT-$\left(k,u\right)$ at time $t$ and frequency $f$ at a specific propagation state can be modeled as \cite{Akdeniz14Millimeter,You17BDMA,Mandar16mmW}
	\begin{align}\label{H_tf_dl}
	\bH_{k,u,v}\left(t,f\right)&=\int\limits_{-1}^{1}\int\limits_{-1}^{1}\int\limits_{-1}^{1}\underbrace{\sqrt{\frac{M_v N_{k,u}}{\zeta_{k,u,v}\left(d_{k,u,v}\right)\eta_{k,u,v}}\varphi_{k,u,v}(\theta,\alpha,\beta)}}_{\triangleq \sqrt{\mathrm{S}_{k,u,v}\left(\theta,\alpha,\beta\right)}}\cdot \mathrm{e}^{\barjmath \psi\left(\theta,\alpha,\beta\right)}
	\cdot \mathrm{e}^{\barjmath 2 \pi \left[t\nu_{k,u,v}\left(\theta\right)-f\tau_{k,u,v}\left(\theta,\alpha,\beta\right)\right]}\ntb
	&\quad\cdot
	\ba_{\mathrm{ut},k,u}\left(\theta\right)\ba_{\mathrm{bs},v}^H\left(\alpha,\beta\right)\intdx{\alpha}\intdx{\beta}\intdx{\theta}\in\bbC^{N_{k,u}\times{M_v}},
	\end{align}
	where $\eta_{k,u,v}$ is the number of channel paths between BS-$v$ and UT-$\left(k,u\right)$. $\psi\left(\theta,\alpha,\beta\right)$ is a random phase uniformly distributed in $\left[0,2\pi\right)$. $\nu_{k,u,v}\left(\theta\right)$ is the Doppler shift associated with $\theta$ and linear over the carrier frequency \cite{You17BDMA}. $\tau_{k,u,v}\left(\theta,\alpha,\beta\right)$ is the path delay associated with auxiliary AoA-AoD pair $\left(\theta,\alpha,\beta\right)$,
	and $\varphi_{k,u,v}\left(\theta,\alpha,\beta\right)$ is the normalized average power of the channel path associated with auxiliary AoA-AoD pair $\left(\theta,\alpha,\beta\right)$ from BS-$v$ to UT-$\left(k,u\right)$. Note that for a specific propagation state, the large-scale path loss $\zeta_{k,u,v}\left(d_{k,u,v}\right)$ can be modeled using \eqref{pathloss}.

	Based on \eqref{H_tf_dl}, we define the DL beam domain channel response matrix from BS-$v$ to UT-$\left(k,u\right)$ at time $t$ and frequency $f$ as \cite{You17BDMA,You15Pilot}
	\begin{align}\label{G_tf_dl}
	\widetilde{\bH}_{k,u,v}\left(t,f\right)\triangleq\bV_{N_{k,u}}^H \bH_{k,u,v}\left(t,f\right)\left(\bV_{M_v^{\mathrm{h}}}\otimes \bV_{M_v^{\mathrm{v}}}\right)\in\bbC^{N_{k,u}\times{M_v}},
	\end{align}
	where $\bV_N\in\bbC^{N\times N}$ is the unitary discrete Fourier transform (DFT) matrix.
    Note that the channel model in \eqref{G_tf_dl} can also be applied to systems with lens antenna array \cite{Zeng16Millimeter}.
	For mmWave/THz massive MIMO channels where $N_{k,u}$, $M_v^{\mathrm{h}}$ and $M_v^{\mathrm{v}}$ are all sufficiently large,
	the beam domain channel matrix elements in \eqref{G_tf_dl} can be well approximated by \cite{You17BDMA}
	\begin{align}\label{G_tf_element_dl}
	\left[\widetilde{\bH}_{k,u,v}\left(t,f\right)\right]_{i,j} \approx& \int\limits_{\alpha_{m_j}}^{\alpha_{m_j+1}}\int\limits_{\beta_{n_j}}^{\beta_{n_j+1}}\int\limits_{\theta_i}^{\theta_i+1}\sqrt{\mathrm{S}_{k,u,v}\left(\theta,\alpha,\beta\right)}\cdot \mathrm{e}^{\barjmath \psi\left(\theta,\alpha,\beta\right)}\ntb
	&\cdot \mathrm{e}^{\barjmath 2 \pi \left[t\nu_{k,u,v}\left(\theta\right)-f\tau_{k,u,v}\left(\theta,\alpha,\beta\right)\right]}\intdx{\alpha}\intdx{\beta}\intdx{\theta},
	\end{align}
	where $m_j\triangleq\langle j \rangle_{M_v^{\mathrm{h}}}$, $n_j\triangleq \left\lfloor j/M_v^{\mathrm{h}}\right\rfloor$,
	$\alpha_{m_j}\triangleq{{2m_j}/{M_v^{\mathrm{h}}}-1}$, $\beta_{n_j}\triangleq{{2n_j}/{M_v^{\mathrm{v}}}-1}$, and  $\theta_{i}\triangleq{{2i}/{N_{k,u}}-1}$.
    As a result, the DL beam domain channel impulse response matrix from BS-$v$ to UT-$\left(k,u\right)$ at time $t$ and delay $\tau$ can be obtained as
	\begin{align}\label{G_t_tau_dl}
	\left[\widetilde{\bH}_{k,u,v}\left(t,\tau\right)\right]_{i,j} =& \int\limits_{\alpha_{m_j}}^{\alpha_{m_j+1}}\int\limits_{\beta_{n_j}}^{\beta_{n_j+1}}\int\limits_{\theta_i}^{\theta_i+1}\sqrt{\mathrm{S}_{k,u,v}\left(\theta,\alpha,\beta\right)}\cdot \mathrm{e}^{\barjmath \psi\left(\theta,\alpha,\beta\right)}\ntb
	&\cdot \mathrm{e}^{\barjmath 2 \pi t \nu_{k,u,v}\left(\theta\right)}\cdot \delta\left(\tau-\tau_{k,u,v}\left(\theta,\alpha,\beta\right)\right)
	\intdx{\alpha}\intdx{\beta}\intdx{\theta},
	\end{align}
    which is the inverse Fourier transform of $\widetilde{\bH}_{k,u,v}\left(t,f\right)$ with respect to $f$.
    Note that the Doppler and delay spreads of the DL channels over a specific receive beam will be much smaller compared with those of the space domain channels due to the directional properties of the mmWave/THz channels \cite{You17BDMA}.
    In the following, we will adopt \eqref{G_t_tau_dl} to simplify the analysis.
		
	\subsection{Transmission Model}
	
	We consider the beam domain transmission for massive MIMO transmission over the mmWave and THz bands. After OFDM modulation, the DL transmitted signal of BS-$v$ in the given OFDM block, $\bx_{v}\left(t\right)\in\bbC^{M_v\times1}$, can be represented as \cite{Hwang09OFDM}
	\begin{align}
	\bx_{v}\left(t\right) = \sum_{s=0}^{N_\mathrm{us}-1}\bx_{v,s}\cdot \mathrm{e}^{\barjmath 2\pi \frac{s}{\dnnot{T}{us}}t},\quad -\dnnot{T}{cp}\le t < \dnnot{T}{us},
	\end{align}
	where $\bx_{v,s}=\sum_{\forall \left(k,u\right)}\bx_{k,u,v,s}$ is the beam domain signal transmitted from BS-$v$ at the $s$-th sub-carrier, and $\bx_{k,u,v,s}$ is the signal from BS-$v$ to UT-$\left(k,u\right)$ at sub-carrier $s$. Here, we omit the OFDM transmission block index for convenience.
	Accordingly, the received beam domain signal by UT-$\left(k,u\right)$ at time $t$ can be expressed as
	\begin{align}\label{y_k,u,dl}
	\by_{k,u}\left(t\right)= \sum_{v=1}^{U} \int\limits_{-\infty}^{\infty}\widetilde{\bH}_{k,u,v}\left(t,\tau\right)\bx_{v}\left(t-\tau\right)\intdx{\tau}\in\bbC^{N_{k,u} \times 1},
	\end{align}
where $\widetilde{\bH}_{k,u,v}\left(t,\tau\right)$ is defined in \eqref{G_t_tau_dl}.
Note that possible inter-block interference (IBI) and noise are ignored in the transmission model \eqref{y_k,u,dl} for clarity.

    The received signal $\by_{k,u}\left(t\right)$ suffers from time offsets $\tau_{k,u}\in\left[\tau_{k,u}^{\mathrm{min}},\tau_{k,u}^{\mathrm{max}}\right]$, and Doppler offsets $\nu_{k,u}\in\left[\nu_{k,u}^{\mathrm{min}},\nu_{k,u}^{\mathrm{max}}\right]$, which severely affect the transmission performance. To mitigate these effects, we perform time and frequency compensation at the UT receiver.
	Different from the traditional space domain synchronization method \cite{Morelli07Synchronization}, we propose to apply the PBS scheme \cite{You17BDMA} in our network transmission framework.
	
	Due to the highly directional propagation properties of the mmWave/THz signals,
	the received signal of a specific UT-$(k,u)$ on a particular receive beam
	 will exhibit smaller delay and Doppler spreads compared with the space domain signals \cite{You17BDMA,Zeng16Millimeter}.
    In other words, the $i$-th element of the beam domain received signal $\by_{k,u}(t)$, i.e., $y_{k,u,i}(t)$, will experience the time offsets ranging from $\tau_{k,u,i}^{\mathrm{min}}$ to $\tau_{k,u,i}^{\mathrm{max}}$ as well as the frequency offsets ranging from $\nu_{k,u,i}^{\mathrm{min}}$ to $\nu_{k,u,i}^{\mathrm{max}}$. We then perform the PBS scheme, i.e., to apply time and frequency adjustment parameters as $\tau_{k,u,i}^{\mathrm{syn}}=\tau_{k,u,i}^{\mathrm{min}}$
	and $\nu_{k,u,i}^{\mathrm{syn}}=(\nu_{k,u,i}^{\mathrm{max}}+\nu_{k,u,i}^{\mathrm{min}})/2$ \cite{You17BDMA}. As a result, the received signal over beam $i$ of UT-$(k,u)$ is given by
	\begin{align}
	y_{k,u,i}^{\mathrm{pbs}}\left(t\right) = y_{k,u,i}(t+\tau_{k,u,i}^{\mathrm{syn}})\cdot \mathrm{e}^{-\barjmath 2\pi\left(t+\tau_{k,u,i}^{\mathrm{syn}}\right)\nu_{k,u,i}^{\mathrm{syn}}}.
	\end{align}
	Then, the received signals over all beams after PBS can be represented as
	\begin{align}
	\by_{k,u}^\mathrm{pbs}\left(t\right) = \left[y_{k,u,0}^{\mathrm{pbs}}\left(t\right)
	,\;y_{k,u,1}^{\mathrm{pbs}}\left(t\right),\;\cdots,\;y_{k,u,N_{k,u}-1}^{\mathrm{pbs}}\left(t\right)\right]^T.
	\end{align}
	
	After CP removal and DFT operation, the demodulated OFDM symbol over beam $i$ of UT-$\left(k,u\right)$ at sub-carrier $s$ is \cite{Hwang09OFDM,You17BDMA}
	\begin{align}
	y_{k,u,i,s} &= \frac{1}{\dnnot{T}{us}}\int_{0}^{\dnnot{T}{us}}y_{k,u,i}^{\mathrm{pbs}}\left(t\right)\cdot \mathrm{e}^{-\barjmath 2 \pi \frac{s}{\dnnot{T}{us}}t}\intdx{t}
	\ntb
	&= \sum_{v=1}^{U}\sum_{j=0}^{M_v-1}\left[{\bG}_{k,u,v,s}\right]_{i,j}\left[\bx_{v,s}\right]_j,
	\end{align}
	where ${\bG}_{k,u,v,s}\in\bbC^{N_{k,u}\times M_v}$ denotes the equivalent beam domain channel from BS-$v$ to UT-$\left(k,u\right)$ after PBS at sub-carrier $s$ in the given block. Its $(i,j)$-th element is given by \cite{You17BDMA}
	\begin{align}\label{G_pbs}
	\left[{\bG}_{k,u,v,s}\right]_{i,j} \triangleq &
	\int\limits_{\alpha_{m_j}}^{\alpha_{m_j+1}}\int\limits_{\beta_{n_j}}^{\beta_{n_j+1}} \int\limits_{\theta_{i}}^{\theta_{i+1}}\sqrt{\mathrm{S}_{k,u,v}\left(\theta,\alpha,\beta\right)}
\cdot \mathrm{e}^{\barjmath \psi\left(\theta,\alpha,\beta\right)}\cdot \mathrm{e}^{-\barjmath 2 \pi \frac{s}{\dnnot{T}{us}} \left(\tau_{k,u,v}\left(\theta,\alpha,\beta\right) - \tau_{k,u,i}^{\mathrm{syn}} \right)}\ntb
	&\cdot \mathrm{e}^{\barjmath 2\pi \tau_{k,u,i}^{\mathrm{syn}}\left(\nu_{k,u,v}\left(\theta\right)-\nu_{k,u,i}^{\mathrm{syn}}\right)}
	\intdx{\alpha}\intdx{\beta}\intdx{\theta} .
	\end{align}
	
	\remark
	In practical OFDM systems, the values of $\dnnot{T}{cp}$ and $\dnnot{T}{us}$ should satisfy the following condition \cite{Hwang09OFDM}
	\begin{align}\label{condition}
	\maxis{k,u}\left\lbrace\Delta_{\tau_{k,u}}\right\rbrace \le \dnnot{T}{cp} \le \dnnot{T}{us} \ll 1/{\maxis{k,u}\left\lbrace\Delta_{\nu_{k,u}}\right\rbrace},
	\end{align}
	where $\Delta_{\tau_{k,u}}$ and $\Delta_{\nu_{k,u}}$ are the effective channel delay and frequency spreads after synchronization, respectively.
    Note that in the traditional space domain synchronization method, the corresponding effective channel delay and Doppler frequency spreads are given by $\Delta_{\tau_{k,u}}^{\mathrm{spa}} = \tau_{k,u}^{\mathrm{max}} - \tau_{k,u}^{\mathrm{min}}$, and $\Delta^{\mathrm{spa}}_{\nu_{k,u}} = \left(\nu_{k,u}^{\mathrm{max}}-\nu_{k,u}^{\mathrm{min}}\right)/2$, respectively \cite{Tse05Fundamentals}.
    As $\Delta_{\nu_{k,u}}^{\mathrm{spa}}$ is linear with the velocity and the carrier frequency \cite{You17BDMA}, it is difficult to select proper OFDM parameters for mmWave/THz transmission in high mobility scenarios. However, with the application of the PBS scheme, the effective delay spread, $\Delta_{\tau_{k,u}}^{\mathrm{per}}= \mathop{\max}\limits_i \left\lbrace\tau_{k,u,i}^{\mathrm{max}} - \tau_{k,u,i}^{\mathrm{min}}\right\rbrace \le \Delta^{\mathrm{spa}}_{\tau_{k,u}}$, and the effective Doppler frequency spread, $\Delta_{\nu_{k,u}}^{\mathrm{per}} = \mathop{\max}\limits_i \left\lbrace\frac{\nu_{k,u,i}^{\mathrm{max}} - \nu_{k,u,i}^{\mathrm{min}}}{2}\right\rbrace =\Delta^{\mathrm{spa}}_{\nu_{k,u}}/N_{k,u}$, can be significantly reduced \cite{You17BDMA}.
    More importantly, the PBS makes the transmission design over mmWave/THz bands approximately the same as that over the regular bands, which can effectively mitigate the mobility issue in mmWave/THz transmission.

	With the PBS scheme applied, we can obtain the network massive MIMO-OFDM beam domain transmission model over each subcarrier given by
	\begin{align}\label{y=gx}
	\by_{k,u,s} = \sum_{v=1}^{U}{\bG}_{k,u,v,s} \bx_{v,s}\in\bbC^{N_{k,u}\times 1},\quad s=0,1,\cdots,N_\mathrm{us} -1.
	\end{align}
    It should be noted that if the traditional synchronization method is adopted, a more complicated transmission model involving, e.g., inter-carrier interference (ICI), should be considered. This will lead to complexity in the transmission design \cite{Hwang09OFDM,You17BDMA}.
	From \eqref{G_pbs}, the elements of the equivalent channel after PBS, ${\bG}_{k,u,v,s}$, are uncorrelated, and the corresponding statistical CSI can be modeled as \cite{You17BDMA}	
\begin{align}
	\bOmega_{k,u,v,s}= \mathbb{E}\left\lbrace{\bG}_{k,u,v,s}\odot {\bG}_{k,u,v,s}^*\right\rbrace,
	\end{align}
	where the element $\left[\bOmega_{k,u,v,s}\right]_{i,j}$ corresponds to the average power of $\left[\bG_{k,u,v,s}\right]_{i,j}$, capturing the average coupling effect between the $j$-th transmit eigenmode of BS-$v$ and the $i$-th receive eigenmode of UT-$\left(k,u\right)$ as
	\begin{align}\label{Omega_dl}
	\left[\bOmega_{k,u,v,s}\right]_{i,j}=\int\limits_{\alpha_{m_j}}^{\alpha_{m_j+1}}\int\limits_{\beta_{n_j}}^{\beta_{n_j+1}} \int\limits_{\theta_{i}}^{\theta_{i+1}}{\mathrm{S}_{k,u,v}\left(\theta,\alpha,\beta\right)}\intdx{\alpha}\intdx{\beta}\intdx{\theta}.
	\end{align}
	Note that the statistical CSI, $\bOmega_{k,u,v,s}$, is independent of sub-carriers; as a result, the sub-carrier subscript $s$ in \eqref{Omega_dl} can be omitted, i.e., $\bOmega_{k,u,v,s} = \bOmega_{k,u,v},\forall s$. Therefore, the transmission strategies utilizing the statistical CSI are identical over all subcarriers. This can significantly reduce the overhead of the CSI acquisition at the transmitter and the complexity in practical wideband transmission strategy design.
	
	\section{Network Downlink Transmission}\label{sec:transmission}

	In this section, we investigate DL transmission with PBS for mmWave/THz network massive MIMO systems. Note that if PBS is not performed as described in \secref{sec:system_model}, it is difficult to select proper OFDM parameters, especially in high mobility scenarios. Also, the ICI as well as IBI should be properly handled. Therefore, we design the network DL transmission strategy in this section based on the PBS scheme and the transmission model in \eqref{y=gx}.

	In our DL network transmission strategy design,
	we assume that only the statistical CSI of UTs in the network is available at the transmitter, e.g., $\bOmega_{k,u,v}$. Based on \eqref{y=gx}, a compact beam domain representation of the network transmission over the mmWave/THz bands can be written as
	\begin{align}\label{y_k_u^dl}
	\by_{k,u} &= \sum_{v=1}^{U}\bG_{k,u,v}\bx_{v} +  \bn_{k,u}\ntb
	&=\sum_{v=1}^{U}\bG_{k,u,v}\bx_{k,u,v} + \sum_{v=1}^{U}\sum_{\left(i,j\right)\neq \left(k,u\right)}\bG_{k,u,v}\bx_{i,j,v} + \bn_{k,u}\ntb
	&=\bG_{k,u}\bx_{k,u} + \sum_{\left(i,j\right)\neq\left(k,u\right)}{\bG}_{k,u}\bx_{i,j} + \bn_{k,u}\in\bbC^{N_{k,u}\times 1},
	\end{align}
	where
	$\bx_{k,u,v}$ is the signal for UT-$\left(k,u\right)$ transmitted from BS-$v$ in the beam domain,
	${\bG}_{k,u}=\left[{\bG}_{k,u,1},{\bG}_{k,u,2},\cdots,{\bG}_{k,u,U}\right]\in\bbC^{N_{k,u}\times M_{\mathrm{tot}}  }$, $\bx_{k,u}=\left[\bx_{k,u,1}^T,\cdots,\bx_{k,u,U}^T\right]^T\in\bbC^{M_{\mathrm{tot}} \times 1}$, and the indices of the OFDM symbols and the subcarriers are omitted for brevity.
	The transmitted signal $\bx_{k,u}$ satisfies $\mathbb{E}\left\lbrace\bx_{k,u}\right\rbrace = \mathbf{0},\forall \left(k,u\right)$, and $\mathbb{E}\lbrace\bx_{k,u}\bx_{i,{j}}^H\rbrace=\mathbf{0},\forall \left(i,j\right)\neq \left(k,u\right)$. The covariance matrix of $\bx_{k,u}$ is $\bQ_{k,u}=\mathbb{E}\lbrace\bx_{k,u}\bx_{k,u}^H\rbrace\in\bbC^{M_{\mathrm{tot}} \times M_{\mathrm{tot}} }$. The noise $\bn_{k,u}\in\bbC^{N_{k,u}\times 1}$ is circularly symmetric complex-valued Gaussian distributed with mean $\mathbf{0}$ and covariance matrix $\sigma^2\bI_{N_{k,u}}$.
	
	Denote by $\bz_{k,u}\triangleq \sum\nolimits_{\left(i,j\right)\neq\left(k,u\right)}{\bG}_{k,u}\bx_{i,j} +\bn_{k,{u}}$ the aggregate noise-plus-interference (NPI) component at UT-$\left(k,u\right)$. With the properly designed DL pilots, each UT-$(k,u)$ is able to obtain its instantaneous CSI \cite{Sun15Beam,You17BDMA}. Using a worst-case transmission design \cite{Hassibi03How}, and assuming the NPI component to be an equivalent Gaussian noise vector, we then have the corresponding covariance matrix $\bK_{k,u}= \mathbb{E}\left\lbrace\bz_{k,u}\bz_{k,u}^H\right\rbrace$ given by
	\begin{align}\label{K^dl}
	\bK_{k,u} = \sigma^2\bI_{N_{k,u}} \!+\! \sum_{\left(i,j\right)\neq\left(k,u\right)}\underbrace{\mathbb{E}\left\lbrace{\bG}_{k,u}\bQ_{i,j}{\bG}_{k,u}^H\right\rbrace}_{\triangleq\bXi_{k,u}\left(\bQ_{i,{j}}\right)} .
	\end{align}
	Note that $\bXi_{k,u}\left(\bX\right)$ defined in \eqref{K^dl} is a matrix-valued function of $\bX$.
	Using the independently distributed  properties of the beam domain channel $\bG_{k,u}$, we can obtain that $\bXi_{k,u}\left(\bX\right)$ is a diagonal matrix with the $n$-th element given by
	\begin{align}\label{Xi_dl}
	\left[\bXi_{k,u}\left(\bX\right)\right]_{n,n} = \tr{\mathsf{diag}\left\lbrace\left(\left[\bOmega_{k,u}\right]_{n,:}\right)^T\right\rbrace\bX},
	\end{align}
	where $\bOmega_{k,u} = \mathbb{E}\lbrace\bG_{k,u} \odot\bG_{k,u}^*\rbrace$ is the beam domain statistical CSI.
	As a result, the DL ergodic transmission rate of UT-$\left(k,u\right)$ is obtained as  \cite{Lu19Robust}
	\begin{align}\label{rate_k_dl}
	R_{k,u}=& \mathbb{E}\left\lbrace\logdet{\bK_{k,u}+{\bG}_{k,u}\bQ_{k,u}{\bG}_{k,u}^H}\right\rbrace- \logdet{\bK_{k,u}}.
	\end{align}

	Eq. \eqref{rate_k_dl} indicates that the rate is a function of $\bQ_{k,u},\forall(k,u)$. Alternatively, we are able to design the network transmission strategy with the statistical CSIT to maximize the network sum-rate under the given constraints. Specifically,
	our design objective is to identify the optimal transmit covariance matrices under a per BS power constraint, which can be formulated as\footnote{Notice that the network transmission design can also be formulated using the weighted sum rate maximization criterion with UTs' quality-of-service requirements taken into account.}
	\begin{align}\label{OP1_dl}
	\argmax{\left\{\bQ_{k,u},\forall \left(k,u\right)\right\}}\quad &R = \sum_{u=1}^U\sum_{k=1}^{K_u}R_{k,u}\ntb
	\mathrm{s.t.}\quad&\tr{\sum_{u=1}^{U}\sum_{k=1}^{K_u}\bE_v \bQ_{k,u}}\le P_v,&\forall v\in\mathcal{U},\ntb
	& \bQ_{k,u} \succeq \mathbf{0},&\forall {\left(k,u\right)}\in\mathcal{K},
	\end{align}
	where $\bE_v = \mathsf{diag}\left\lbrace \bzero_{1\times \sum_{v'=1}^{v-1}M_{v'}  }, \; \mathbf{1}_{1\times M_v}, \; \bzero_{1\times \sum_{v'=v+1}^{U}M_{v'}  } \right\rbrace$, and $P_v$ denotes the power budget of BS-$v$.

	\subsection{Optimality of Beam Domain Transmission}
	
	We denote the eigenvalue decomposition of the beam domain transmit covariance matrix as $\bQ_{k,u}=\mathbf{U}_{k,u}\bLambda_{k,u}\mathbf{U}_{k,u}^H$. $\mathbf{U}_{k,u}$ represents the subspace in which the transmitted signals from all BSs to UT-$\left(k,u\right)$ fall, and the elements of the diagonal matrix $\bLambda_{k,u}$ represent the power allocated to each direction of the subspace for the transmit signals.
	The following theorem identifies the optimal eigenmatrix $\mathbf{U}_{k,u},\forall(k,u)$.
	\begin{theorem}\label{prop_optimal_dl}
		For problem \eqref{OP1_dl}, the eigenmatrix, $\mathbf{U}_{k,u}, \forall (k,u)$, of the optimal transmit covariance matrix, $\bQ_{k,u},\forall (k,u)$, exhibits the following structure
		\begin{align}
		\mathbf{U}_{k,u} = \bI_{M_{\mathrm{tot}}},\quad\forall (k,u).
		\end{align}
	\end{theorem}
	\begin{IEEEproof}
		Please refer to \appref{optimal_dl}.
	\end{IEEEproof}
	
	\thref{prop_optimal_dl} reveals that to maximize the DL network sum-rate $R$, the signal transmission can be performed in the beam domain.
	In addition, according to \thref{prop_optimal_dl}, $\mathbb{E}\lbrace\bx_{k,u,v}\bx_{k,u,v'}^H\rbrace=\mathbf{0}\;\left(\forall v\neq v'\right)$, implying that it is not necessary to jointly design the signals sent by different BSs. In other words, different BSs can work individually when sending signals to a specific UT.

	Based on \thref{prop_optimal_dl}, the optimization of the transmit covariance matrices in \eqref{OP1_dl} can be simplified as the following beam domain power allocation problem, given by
	\begin{align}\label{OP2_dl}
	\argmax{\bLambda\triangleq\left\lbrace\bLambda_{k,u},\forall \left(k,u\right)\right\rbrace}\quad &\sum_{u=1}^U\sum_{k=1}^{K_u}\left(f_{k,u}^+\left(\bLambda\right) - f_{k,u}^-\left(\bLambda\right)\right)\ntb
	\mathrm{s.t.}\quad&\tr{\sum_{u=1}^{U}\sum_{k=1}^{K_u}\bE_v \bLambda_{k,u}}\le P_v,\quad\forall v\in\mathcal{U},\ntb
	& \bLambda_{k,u} \succeq \mathbf{0},\; \bLambda_{k,u}\; \mathrm{diagonal},\qquad\forall \left(k,u\right)\in\mathcal{K},
	\end{align}
	where
	\begin{align}
	f_{k,u}^+\left(\bLambda\right)&\!\triangleq\!\mathbb{E}\left\lbrace\logdet{\bK_{k,u}\left(\bLambda\right)+{\bG}_{k,u}\bLambda_{k,u}{\bG}_{k,u}^H}\right\rbrace \label{f_k_u^+},  \\
	f_{k,u}^-\left(\bLambda\right)&\!\triangleq\!\logdet{\bK_{k,u}\left(\bLambda\right)}\label{f_k_u^-},	\\
	\bK_{k,u}\left(\bLambda\right)&\!\triangleq\!\sigma^2\bI_{N_{k,u}}+\sum_{\left(i,j\right)\neq \left(k,u\right)}\mathbb{E}\left\lbrace{\bG}_{k,u}\bLambda_{i,j}{\bG}_{k,u}^H\right\rbrace.
	\end{align}
	Compared to the original transmit covariance optimization problem in \eqref{OP1_dl}, the number of the optimization variables in \eqref{OP2_dl} is significantly reduced.
	In the following, we will develop efficient algorithms to solve the problem in \eqref{OP2_dl}.
	
	\subsection{Iterative Algorithm for DL Power Allocation}
	Now we focus on the design of the DL power allocation algorithm. We first utilize CCCP to find a local optimum of problem \eqref{OP2_dl}. In order to reduce the computational complexity, we then develop a DE based power allocation algorithm.
	
	\subsubsection{CCCP Method}

    Since $f_{k,u}^+\left(\bLambda\right)$ in \eqref{f_k_u^+} and $f_{k,u}^-\left(\bLambda\right)$ in \eqref{f_k_u^-} are both concave, the objective in \eqref{OP2_dl} is a difference of concave (d.c.) functions, and the problem is known to be non-deterministic polynomial time hard (NP-hard).
    For the d.c. problem, it has been shown that the CCCP approach, i.e., a minorization-maximization algorithm, can be an efficient solution \cite{Sun17Majorization}.
	The main idea of CCCP is to transform a non-convex problem to a series of easy-to-handle subproblems \cite{Sun17Majorization}. In particular, we replace $f_{k,u}^-\left(\bLambda\right)$ with its first-order Taylor expansion and then solve the subproblem in each CCCP iteration, which further yields the next iteration.
	Then, the problem in \eqref{OP2_dl} is tackled through solving the following sequence of optimization subproblems iteratively
	\begin{align}\label{OP3_dl}
	\bLambda^{\left(\ell+1\right)} =\argmax{\bLambda}\quad&\sum_{u=1}^{U}\sum_{k=1}^{K_u}\left\lbrace f_{k,u}^+\left(\bLambda\right)-\widetilde{f}_{k,u}^-\left(\bLambda;\bLambda^{\left(\ell\right)}\right)\right\rbrace\ntb
	\mathrm{s.t.}\quad &\mathrm{constraints \quad in \quad \eqref{OP2_dl}},
	\end{align}
	where
	\begin{align}
	\widetilde{f}_{k,u}^-\left(\bLambda;\bLambda^{\left(\ell\right)}\right)=& f_{k,u}^-\left(\bLambda^{\left(\ell\right)}\right)+ \mathrm{tr}\left((\boldsymbol{\Delta}_{k,u}^{(\ell)})^T\left(\bLambda_{k,u}-\bLambda_{k,u}^{\left(\ell\right)}\right)\right).
	\end{align}
In the $\ell$-th iteration, $\boldsymbol{\Delta}_{k,u}^{\left(\ell\right)}=\frac{\partial}{\partial \bLambda_{k,u}}\sum_{i=1}^U\sum_{j=1}^{K_i}f_{i,j}^-\left(\bLambda^{\left(\ell\right)}\right)$ is a diagonal matrix, which can be expressed as
	\begin{align}\label{Delta}
	\boldsymbol{\Delta}_{k,u}^{\left(\ell\right)} = \sum_{\left(i,j\right)\neq \left(k,u\right)}\sum_{n=1}^{N_{i,j}}\frac{\hatbR_{i,j,n}}{\sigma^2+\mathrm{tr}\left(\bLambda_{\backslash{\left(i,j\right)}}^{\left(\ell\right)}\hatbR_{i,j,n}\right)},
	\end{align}
	where $\bLambda_{\backslash{\left(i,j\right)}}^{\left(\ell\right)}\!=\!\sum\nolimits_{\left(p,q\right)\neq \left(i,j\right) }\bLambda_{p,q}^{\left(\ell\right)}$, and $\hatbR_{i,j,n}=\diag{\left[\bOmega_{i,j}\right]_{n,:}}$.
	Note that the $t$-th diagonal element of $\boldsymbol{\Delta}_{k,u}^{\left(\ell\right)}$ is given by
	\begin{align}
	\left[\boldsymbol{\Delta}_{k,u}^{\left(\ell\right)}\right]_{t,t}=\sum_{\left(i,j\right)\neq\left(k,u\right)}\sum_{n=1}^{N_{i,j}}\frac{\left[\bOmega_{i,j}\right]_{n,t}}{\sigma^2+\sum\limits_{\left(p,q\right)\neq \left(i,j\right)}\sum\limits_{m=1}^{M_{\mathrm{tot}}}\left[\bLambda_{p,q}^{\left(\ell\right)}\right]_{m,m}\left[\bOmega_{i,j}\right]_{n,m}}.
	\end{align}
	
    According to \cite{Sun17MulcellBeam}, the objective value sequence generated by \eqref{OP3_dl} will converge. Note that CCCP is an efficient method to solve the d.c. problem, and the optimization result is a locally optimal solution of the original problem in \eqref{OP2_dl}.

	\subsubsection{DE Method}
	Note that it is challenging to derive closed-form expressions of $f_{k,u}^+\left(\bLambda\right)$ due to the expectation operation.
	If the Monte-Carlo method is adopted, the optimization can be computationally cumbersome as sample averaging is required.
	Via utilizing the large dimensional random matrix theory \cite{Lu16Free,Couillet11Random}, the ergodic rate expression can be well approximated by its DE. In particular, the DE of $f_{k,u}^+\left(\bLambda\right)$ is given by
	\begin{align}\label{f_k_u_de^+^dl}
	\overline{f}_{k,u}^+\left(\bLambda\right) \!=\! \logdet{\bI_{M_\mathrm{tot}} \!+\! \mathbf{\Gamma}_{k,u}\bLambda_{k,u}} \!+\! \logdet{\widetilde{\mathbf{\Gamma}}_{k,u} \!+\! \bK_{k,u}\left(\bLambda\right)} \!-\!
	\mathrm{tr}\left(\bI_{N_{k,u}}
	\!-\! \widetilde{\bPhi}_{k,u}^{-1}\right),
	\end{align}
	where $\mathbf{\Gamma}_{k,u}$ and $\widetilde{\mathbf{\Gamma}}_{k,u}$ are given by
	\begin{align}
	\mathbf{\Gamma}_{k,u}&=\mathbf{\Pi}_{k,u}\left(\widetilde{\bPhi}_{k,u}^{-1}\bK_{k,u}^{-1}\right)\in\bbC^{M_{\mathrm{tot}}\times M_{\mathrm{tot}}}, \label{Gamma_dl}\\
	\widetilde{\mathbf{\Gamma}}_{k,u}&=\bXi_{k,u}\left(\bPhi_{k,u}^{-1}\bLambda_{k,u}\right)\in\bbC^{N_{k,u}\times N_{k,u}},\label{Gamma_tilde_dl}
	\end{align}
	respectively.
	$\widetilde{\bPhi}_{k,u}$ and $\bPhi_{k,u}$ are obtained by the following iterative equations
	\begin{align}
	\widetilde{\bPhi}_{k,u}&=\bI_{N_{k,u}} + \bXi_{k,u}\left(\bPhi_{k,u}^{-1}\bLambda_{k,u}\right)\bK_{k,u}^{-1},\label{Phi_tilde_dl}\\
	\bPhi_{k,u} &= \bI_{M_\mathrm{tot}} +\mathbf{\Pi}_{k,u}\left(\widetilde{\bPhi}_{k,u}^{-1}\bK_{k,u}^{-1}\right)\bLambda_{k,u}.\label{Phi_dl}
	\end{align}
	Note that
	$\mathbf{\Pi}_{k,u}\left(\bY\right)\triangleq\mathbb{E}\lbrace{\bG}_{k,u}^H\bY{\bG}_{k,u}\rbrace$ above is a diagonal matrix-valued function with its $m$-th element being
	\begin{align}
	\left[\bPi_{k,u}\left(\bY\right)\right]_{m,m} &= \tr{\mathsf{diag}\left\lbrace\left[\bOmega_{k,u}\right]_{:,m}\right\rbrace\bY}.
	\end{align}
	
	By replacing $f_{k,u}^+\left(\bLambda\right)$ with its DE $\overline{f}_{k,u}^+\left(\bLambda\right)$ in \eqref{f_k_u_de^+^dl} in each iteration, the following series of problems instead of \eqref{OP3_dl} are considered
	\begin{align}\label{OP4_dl}
	\bLambda^{\left(\ell+1\right)} =\argmax{\bLambda}\quad&\sum_{u=1}^{U}\sum_{k=1}^{K_u}\left\lbrace \overline{f}_{k,u}^+\left(\bLambda\right)-\widetilde{f}_{k,u}^-\left(\bLambda;\bLambda^{\left(\ell\right)}\right)\right\rbrace\ntb
	\mathrm{s.t.}\quad &\mathrm{constraints \quad in \quad \eqref{OP2_dl}}.
	\end{align}
Compared with the Monte-Carlo method that averages over the channel realizations for expectation operation, the DE expression $\overline{f}_{k,u}^+\left(\bLambda\right)$ can be calculated using the statistical CSI, i.e., $\bOmega_{k,u}, \forall(k,u)$, in a few iterations with high accuracy \cite{Lu16Free}. In addition, $\overline{f}_{k,u}^+\left(\bLambda\right)$ is strictly concave on $\bLambda$ \cite{Dumont10DEconcave}. Therefore, each sub-problem in \eqref{OP4_dl} is concave with respect to $\bLambda$, and the resulting solution sequence is still guaranteed to converge.
	
	For notational clarity, we define the DE of the DL network transmission sum-rate in the $\ell$-th iteration as
	\begin{align}\label{sum_rate_de_dl}
	\overline{R}^{\left(\ell\right)}=\sum_{u=1}^{U}
	\sum_{k=1}^{K_u}\left\lbrace\overline{f}_{k,u}^+\left(\bLambda^{\left(\ell\right)}\right)-f_{k,u}^-\left(\bLambda^{\left(\ell\right)}\right)\right\rbrace.
	\end{align}
	We use set $\mathcal{C}_{v}= \left\lbrace m | \sum_{v'=1}^{v-1}M_{v'}+1 \le m \le \sum_{v'=1}^{v}M_{v'}\right\rbrace$ to represent the indices of the elements in $\bLambda_{k,u}$ that are related to BS-$v$. We denote the $m$-th diagonal entries of $\mathbf{\Gamma}_{k,u}$, $\widetilde{\mathbf{\Gamma}}_{k,u}$, $\bLambda_{k,u}$, $\boldsymbol{\Delta}_{k,u}$, and $\hatbR_{i,j,n}$ as $\mathbf{\gamma}_{k,u,m}$, $\widetilde{\gamma}_{k,u,m}$, $\lambda_{k,u,m}$, $\delta_{k,u,m}$, and $\hat{r}_{i,j,m,n}$, respectively.
	For problem \eqref{OP4_dl} in the $\ell$-th CCCP iteration, we have the following theorem.
	
	\begin{theorem}\label{prop_Lambda_dl}
		The solution to problem \eqref{OP4_dl} is equivalent to that to the following problem
		\begin{align}\label{OP5_dl}
		\max_{\bLambda}\quad&\sum_{u=1}^{U}\sum_{k=1}^{K_u}\left\lbrace \logdet{\bI_{M_\mathrm{tot}}+\mathbf{\Gamma}_{k,u}\bLambda_{k,u}}
		+\logdet{\widetilde{\mathbf{\Gamma}}_{k,u}+\bK_{k,u}\left(\bLambda\right)}-\mathrm{tr}\left(\boldsymbol{\Delta}_{k,u}^{\left(\ell\right)}\bLambda_{k,u}\right)\right\rbrace\ntb
		\mathrm{s.t.}\quad&\mathrm{constraints \quad in \quad \eqref{OP2_dl}}.
		\end{align}
		For a given BS-$v$, the $m$-th element $\lambda_{k,u,m}^{\left(\ell+1\right)}$ of $\bLambda_{k,u}^{\left(\ell+1\right)}$, $m\in \mathcal{C}_v$, can be calculated via solving the following equations
		\begin{align}\label{lambda_dl}
		\left\{
			{\begin{array}{*{20}{c}}
			{
			\frac{\gamma_{k,u,m}^{\left(\ell\right)}}{1+\gamma_{k,u,m}^{\left(\ell\right)}\lambda_{k,u,m}^{\left(\ell+1\right)}} +\sum\limits_{\left(i,j\right)\neq \left(k,u\right)}\sum\limits_{n=1}^{N_{i,j}}\frac{\hat{r}_{i,j,m,n}}{\widetilde{\gamma}_{i,j,n}^{\left(\ell\right)}+\sigma^2+\mathrm{tr}\left(\hatbR_{i,j,n}\bLambda_{\backslash \left(i,j\right)}^{\left(\ell+1\right)}\right)}=\delta_{k,u,m}^{\left(\ell\right)}+\mu_v,
			\mu_v<\chi_{k,u,m}^{\left(\ell+1\right)}-\delta_{k,u,m}^{\left(\ell\right)},
			}\\
			{
			\lambda_{k,u,m}^{\left(\ell+1\right)}=0,\qquad\qquad\qquad\qquad\qquad\qquad\qquad\qquad\qquad\qquad\qquad
			\mu_v\ge\chi_{k,u,m}^{\left(\ell+1\right)}-\delta_{k,u,m}^{\left(\ell\right)}
		    }.
			\end{array}}
			\right.
		\end{align}
		The Lagrange multipliers, $\mu_v,v\in\mathcal{U}$, satisfy the following KKT conditions
		\begin{align}\label{mu}
		\mu_v\left(\mathrm{tr}\left(\sum_{u=1}^{U}\sum_{k=1}^{K_u}\bE_v\bLambda_{k,u}\right) - P_v\right) = 0, \ntb
		\mu_v\ge 0,
		\end{align}
		and the auxiliary variable $\chi_{k,u,m}^{\left(\ell+1\right)}$ is given by
		\begin{align}\label{chi_dl}
		\chi_{k,u,m}^{\left(\ell+1\right)}=	\gamma_{k,u,m}^{\left(\ell+1\right)}+ \sum\limits_{\left(i,j\right)\neq \left(k,u\right)}\sum\limits_{n=1}^{N_{i,j}}\frac{\hat{r}_{i,j,m,n}}{\widetilde{\gamma}_{i,j,n}^{\left(\ell+1\right)} +\sigma^2+ \sum\limits_{\left(p,q,m'\right) \in \mathcal{S}_{k,u,m,i,j}} \hat{r}_{i,j,m',n}\lambda_{p,q,m'}^{\left(\ell+1\right)}},
		\end{align}
		where $\mathcal{S}_{k,u,m,i,j}$ is the set given by
		\begin{align}
		\mathcal{S}_{k,u,m,i,j}=\left\lbrace\left(p,q,m'\right)|\left(p,q\right)\neq \left(i,j\right),\left(p,q,m'\right)\neq\left(k,u,m\right),p_q\in\mathcal{K},m\in\left\lbrace1,2,\dots,M_\mathrm{tot}\right\rbrace\right\rbrace.
		\end{align}

	\end{theorem}
	
	\begin{IEEEproof}
		Please refer to \appref{Lambda_dl}.
	\end{IEEEproof}
	
    We detail the developed low-complexity beam domain power allocation algorithm in \textbf{Algorithm 1}. Note that to obtain $\bLambda^{\left(\ell+1\right)}$ in Step $12$ of \textbf{Algorithm 1}, an iterative water-filling \textbf{Algorithm 2} is utilized via exploiting \thref{prop_Lambda_dl}, where the auxiliary variables adopted in \textbf{Algorithm 2} are defined as
	\begin{align}
	\rho_{k,u,m}^{\left(\ell\right)}\left(x_{k,u,m}\right)
&=\sum\limits_{\left(i,j\right)\neq\left(k,u\right)}\sum\limits_{n=1}^{N_{i,j}}\frac{\hat{r}_{i,j,m,n}}{\widetilde{\gamma}_{i,j,n}^{\left(\ell\right)} +\sigma^2+ \hat{r}_{i,j,m,n}x_{k,u,m} + \sum\limits_{\left(p,q,m'\right) \in \mathcal{S}_{k,u,m,i,j}} \hat{r}_{i,j,m',n}x_{p,q,m'}}
\ntb
	&\qquad+\frac{\gamma_{k,u,m}^{\left(\ell\right)}}{1+\gamma_{k,u,m}^{\left(\ell\right)}x_{k,u,m}}-\delta_{k,u,m}^{\left(\ell\right)}-\mu_v\label{rho},\\
	{\rho'}_{k,u,m}^{\left(\ell\right)}\left(x_{k,u,m}\right)
&=\sum\limits_{\left(i,j\right)\neq\left(k,u\right)}\sum\limits_{n=1}^{N_{i,j}}\frac{-\hat{r}_{i,j,m,n}^2}{\left(\widetilde{\gamma}_{i,j,n}^{\left(\ell\right)} +\sigma^2+ \hat{r}_{i,j,m,n}x_{k,u,m} +  \sum\limits_{\left(p,q,m'\right) \in \mathcal{S}_{k,u,m,i,j}} \hat{r}_{i,j,m',n}x_{p,q,m'}\right)^2}\ntb
	&\qquad-\frac{\left(\gamma_{k,u,m}^{\left(\ell\right)}\right)^2}{\left(1+\gamma_{k,u,m}^{\left(\ell\right)}x_{k,u,m}\right)^2}\label{rhoo},\\
	\mu_v^{\mathrm{max}}&=\maxis{k,u,m}\left\lbrace\gamma_{k,u,m}^{\left(\ell\right)} + \sum_{\left(i,j\right)\neq \left(k,u\right)}\sum_{n=1}^{N_{i,j}}\frac{\hat{r}_{i,j,m,n}}{\widetilde{\gamma}_{i,j,n}^{\left(\ell\right)}+\sigma^2} - \delta_{k,u,m}^{\left(\ell\right)}\right\rbrace \label{u_max},
	\end{align}
	respectively.
	\begin{algorithm}[!t]
		\caption{DE based DL Iterative Power Allocation Algorithm}
		\label{algorithm1}
		\begin{algorithmic}[1]
			\Require Initial power allocation $\bLambda^{(0)}$
			\State Initialize iteration index $\ell = 0$ and calculate $\overline{R}^{\left(\ell\right)}$ as \eqref{sum_rate_de_dl}.
			\Repeat
			\For {all {$ k_u \in \mathcal{K}$}}
			\State Initialize $t=0$ and $\widetilde{\bPhi}_{k,u}^{\left(t\right)}$.
			\Repeat
			\State Calculate $\widetilde{\bPhi}_{k,u}^{\left(t+1\right)}$ and $\bPhi_{k,u}^{\left(t+1\right)}$ by \eqref{Phi_tilde_dl} and \eqref{Phi_dl}.
			\State $t = t+1$.
			\Until convergence of $\widetilde{\bPhi}_{k,u}^{\left(t\right)}$
			\State Calculate $\mathbf{\Gamma}_{k,u}^{\left(\ell\right)}$ and $\widetilde{\mathbf{\Gamma}}_{k,u}^{\left(\ell\right)}$ by \eqref{Gamma_dl} and \eqref{Gamma_tilde_dl}.
			\State Calculate $\boldsymbol{\Delta}_{k,u}^{\left(\ell\right)}$ based on \eqref{Delta}.
			\EndFor
			
			\State Solve \eqref{OP4_dl} to obtain $\bLambda^{\left(\ell+1\right)}$ by \textbf{Algorithm 2}.
			\State Set $\ell = \ell+1$ and calculate $\overline{R}^{\left(\ell\right)}$ by \eqref{sum_rate_de_dl}.
			
			\Until convergence of $\overline{R}^{\left(\ell\right)}$	
			\Ensure Power allocation $\bLambda$		
		\end{algorithmic}
	\end{algorithm}
	\begin{algorithm}
		\caption{DL Iterative Water-filling Algorithm}
		\label{algorithm2}
		\begin{algorithmic}[1]
			\Require Initial power allocation $\bLambda^{(\ell)}$ and transmit power $P_v,\forall v$
			\State Initialize diagonal matrices $\overline{\bX}_{k_,u}^{\left(0\right)}=\bLambda_{k,u}^{\left(\ell\right)},\forall \left(k,u\right) \in\mathcal{K}$, threshold $\epsilon$, and $\overline{x}_{k,u,m}$ is the $m$-th diagonal entries of $\overline{\bX}_{k,u}$.
			\For {all $v \in\mathcal{U}$}
			\State Set iteration index $j=0$, $\mu^{\mathrm{min},j}_v=0$. Calculate $\mu^{\mathrm{max},j}_v$ by \eqref{u_max}.
			\State Initialize $\mu_v^{j} = \left(\mu^{\mathrm{min},j}_v + \mu^{\mathrm{max},j}_v \right)/2 $.
			\Repeat
			\For {all $\left(k,u\right) \in \mathcal{K}$}
			\For {$  m \in \mathcal{C}_v $}
			\State Set $w=0$ and $x_{k,u,m}^{\left( j,w\right)}=\overline{x}_{k,u,m}^{\left(j\right)}$.
			\Repeat
			\State Calculate $\rho_{k,u,m}^{\left(\ell\right)}\left({x}_{k,u,m}^{\left( j,w\right)}\right)$ and $\rho_{k,u,m}^{'\left(\ell\right)}\left(x_{k,u,m}^{\left( j,w\right)}\right)$ as \eqref{rho} and \eqref{rhoo}.
			\State Update $x_{k,u,m}^{\left( j,\left(w+1\right)\right)} = x_{k,u,m}^{\left( j,{w}\right)} - \rho_{k,u,m}^{\left(\ell\right)}\left({x}_{k,u,m}^{\left( j,w\right)}\right)/\rho_{k,u,m}^{'\left(\ell\right)}\left({x}_{k,u,m}^{\left( j,w\right)}\right)$.
			\State  Set $w = w+1$.
			\Until convergence of $x_{k,m}^{\left( j,w\right)}$
			\State Update $\overline{x}_{k,u,m}^{\left(j\right)} = [x_{k,u,m}^{\left( j,w\right)}]^+$.
			\EndFor
			\EndFor
			\State Calculate $p^{\mathrm{tot}}_v=\sum\limits_{\left(k,u\right)\in\mathcal{K}}\sum\limits_{m\in\mathcal{C}_v}\overline{x}^{\left(j\right)}_{k,u,m}$.
			\If {$p^{\mathrm{tot}}_v\le P_v$}
			\State Update $\mu_v^{\mathrm{max},j+1} = \mu_v^{j}$ and $\mu_v^{\mathrm{min},j+1} = \mu_v^{\mathrm{min},j}$.
			\Else
			\State Update $\mu_v^{\mathrm{min},j+1} = \mu_v^{j}$ and $\mu_v^{\mathrm{max},j+1} = \mu_v^{\mathrm{max},j}$.
			\EndIf
			\State Update $\mu_v^{j+1} = \left(\mu^{\mathrm{min},j+1}_v + \mu^{\mathrm{max},j+1}_v \right)/2 $ and set $j=j+1$.
			\Until $\left|p^{\mathrm{tot}}_v-P_v\right|\le\epsilon$
			\EndFor
			\State Update $\bLambda_{k,u}^{\left(\ell+1\right)}=\overline{\bX}_{k,u},\;\forall \left(k,u\right) \in \mathcal{K}$.
			\Ensure Power allocation $\bLambda^{(\ell+1)}$
		\end{algorithmic}
	\end{algorithm}
	\remark
	The solution to \eqref{lambda_dl} has the form of the classical water-filling structure, and the water level depends on the Lagrange multiplier $\mu_v$. For our considered multi-UT scenario, solving \eqref{lambda_dl} is challenging because of the summation of the fractional functions. Newton-Raphson method \cite{Cormen09introduction} is utilized to find the approximate roots of \eqref{lambda_dl} in Step $11$ of \textbf{Algorithm 2}. In addition, to find the optimal Lagrange multipliers, $\mu_v,\forall v\in\mathcal{U}$, the bisection method is utilized. Note that for the case with a single UT-$\left(k,u\right)$, if the power constrains, $\mathrm{tr}\left(\bE_v\bLambda_{k,u}\right)=P_v,\forall v$, are considered, the solution can be obtained in closed-form as follows
	\begin{align}\label{Single_UT}
	\lambda_{k,u,m}^{\left(\ell+1\right)}=&\left[\left(\delta_{k,u,m}^{\left(\ell\right)}+\mu_v\right)^{-1}-\left(\gamma_{k,u,m}^{\left(\ell+1\right)}\right)^{-1}\right]^+,\quad m\in\mathcal{C}_v,\quad \forall v,
	\end{align}
	where $\mu_v$ is chosen to satisfy the constraint 	$\mathrm{tr}\left(\bE_v\bLambda_{k,u}\right) = P_v$.
	
	\section{Numerical Results}\label{sec:simulation}
    In this section, numerical results are presented to evaluate the performance of the proposed network massive MIMO transmission over the mmWave/THz bands, where two typical carrier frequencies, $28$ GHz and $100$ GHz, are adopted.
	We consider a $120^\circ$ three-sector cellular model \cite{Huang12network}, where the cell radius is set to be $100$ meters, and the UTs are uniformly distributed in each sector.
	\tabref{simulation_set_up} illustrates the simulation setup.
    We utilize the mmWave/THz channel model similar to that in \cite{Akdeniz14Millimeter,You17BDMA},
    where the number of channel clusters is set to $4$, and each cluster consists of $20$ subpaths. The delay spread and angle spread are set to be $1388.4$ ns and $2^\circ$, respectively \cite{Akdeniz14Millimeter}. In addition, the large-scale fading parameters are summarized in \tabref{parameter} according to \cite{Rappaport13Millimeter}.
	
	\newcolumntype{L}{>{\hspace*{-\tabcolsep}}l}
	\newcolumntype{R}{c<{\hspace*{-\tabcolsep}}}
	\definecolor{lightblue}{rgb}{0.93,0.95,1.0}
	\begin{table}[!t]
		\caption{Simulation Setup Parameters}\label{simulation_set_up}
		\centering
\footnotesize
		\ra{1.2}
	\begin{tabular}{LR}
		\toprule
		Parameter &   {Value}\\
		\midrule
		\rowcolor{lightblue}
		Carrier frequency &  {$28$ GHz, $100$ GHz}\\
		
		Sampling interval $\dnnot{T}{s}$ & {$6.51$ ns}\\

		\rowcolor{lightblue}
		Subcarrier spacing & {$75$ kHz}\\
		
		Number of subcarriers $\dnnot{N}{us}$ & {2048}\\
		
		\rowcolor{lightblue}
		CP length $\dnnot{N}{cp}$ & {144}\\
		
		Cell radius $R$ & {$100$ m, $\forall u$} \\
		
		\rowcolor{lightblue}
		Number of cells $U$   &{$3$} \\
		
		BS array topology   & {UPA with half-wavelength antenna spacing} \\
		
		\rowcolor{lightblue}
		Number of BS antennas $M_u^{\mathrm{h}} \times M_u^{\mathrm{v}}$  &{$32 \times 4$, $\forall u$} \\
		
		Number of UTs in each sector $K_u$ & {$4,\forall{u}$} \\
		
		\rowcolor{lightblue}
		UT array topology  &  {ULA with half-wavelength antenna spacing}\\
		
		Number of UT antennas $N_{k,u}$  & {$16,\ 64, \forall {\left(k,u\right)}$}\\
		
		\rowcolor{lightblue}
		Noise variance $\sigma^2$  &{$-40$ dBm} \\			
		\bottomrule
	\end{tabular}
	\end{table}
	

\begin{table}[htbp]
	\caption{Large-Scale Fading Parameters}\label{parameter}
\begin{center}
	\begin{tabular}{|m{2.3cm}|m{1.2cm}|m{2.5cm}|m{2.7cm}|m{2.7cm}|}
		\hline
		Parameters & Model  & LOS  & NLOS  & outage  \\ [3pt]\hline
		\multirow{3}{*}{Path loss} & \multirow{3}{*}{Eq. \eqref{pathloss}} & $\mathrm{a}^{\mathrm{LOS}} = 61.4$ & $\mathrm{a}^{\mathrm{NLOS}} = 72.0$ &  \multirow{3}{*}{$\zeta^{\mathrm{out}}_{k,u,v}=\infty$ \cite{Akdeniz14Millimeter}} \\ [3pt]\cline{3-4}
		& &  $\mathrm{b}^{\mathrm{LOS}}=2$ & $\mathrm{b}^{\mathrm{NLOS}}=2.92$ &                   \\ [3pt] \cline{3-4}
		& & $\xi^{\mathrm{LOS}} = 5.8$ dB & $\xi^{\mathrm{NLOS}} = 8.7$ dB &                   \\[3pt] \hline
		Channel state & Eq. \eqref{out_LOS_NLOS}                & \multicolumn{3}{l|}{$1/a_{\mathrm{out}}=30.0$ m, $1/a_{\mathrm{los}} = 67.1$ m, $b_{\mathrm{out}}=5.2$}   \\ [3pt]\hline
	\end{tabular}
\end{center}
\end{table}

	We consider the following two approaches as the performance comparison baseline: 1) the \textbf{coordinated} transmission approach, where the BSs utilize the statistical CSI of all UTs in the network but only transmit signals to the UTs in its cell \cite{Sun17MulcellBeam}; 2) the traditional \textbf{single-cell} transmission approach, where the BS in each cell utilizes the statistical CSI of UTs in its cell and transmits signals to the corresponding UTs.		
		
\begin{figure}
		\centering
		\includegraphics[width=10cm]{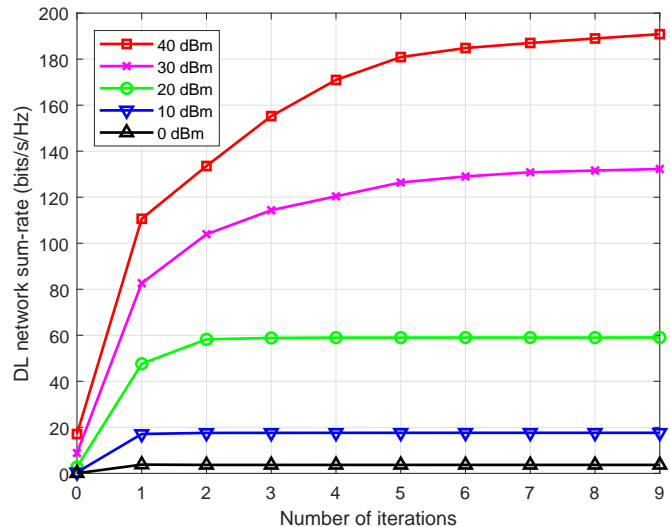}
		\caption{Convergence behavior of \textbf{Algorithm 1}. Results are shown versus the numbers of iterations for different values of transmit power with a carrier frequency of $28$ GHz.}
		\label{iter}
	\end{figure}

	\figref{iter} presents the convergence behavior of \textbf{Algorithm 1} under different values of transmit power budgets. The simulation results indicate that our proposed \textbf{Algorithm 1} generates a non-decreasing DL network sum-rate sequence and converges fast in typical transmit power budget regions. In particular, the DL network sum-rate can usually converge after only one or two iterations in the case of low transmit power budgets.
	
	\begin{figure}
		\centering
		\includegraphics[width=10cm]{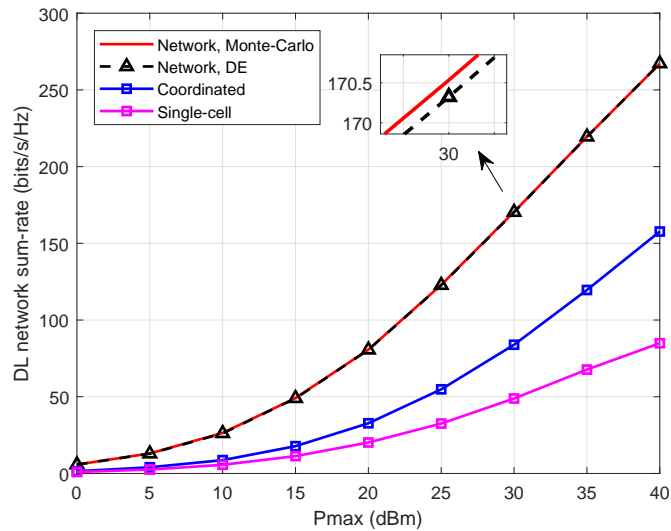}
		\caption{The DL network sum-rate performance comparison between the proposed network, coordinated, and single-cell transmission approaches with a carrier frequency of $28$ GHz. The DE results are also depicted.}
		\label{compare}
	\end{figure}

	\figref{compare} compares the DL network sum-rate performance of the proposed network transmission approach with the baseline ones. We observe that the proposed network transmission approach outperforms the coordinated and single-cell baseline ones, especially at the high power budget regime. Notably, for the case with $P_{\mathrm{max}} = 40$ dBm, the proposed network transmission approach can provide about $70 \%$ performance gains over the coordinated ones. In addition, we can also observe that the adopted DE results are almost identical to those obtained from the Monte-Carlo method.
	
	\begin{figure}
		\centering
		\includegraphics[width=10cm]{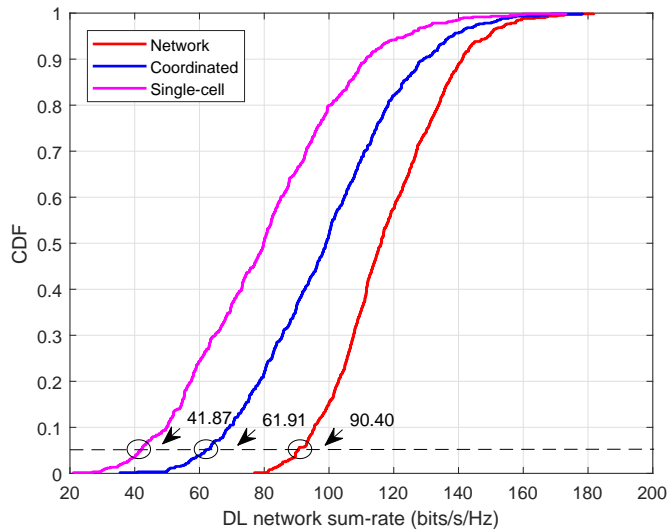}
		\caption{Cumulative distributive function (CDF) performance of the DL network sum-rate for different approaches with $28$ GHz carrier frequency, $M_u = 16\times 4,\forall u, N_{k,u} =8,\forall \left(k,u\right)\in\mathcal{K}, P_{\mathrm{max}}=40$ dBm.}
		\label{cdf}
	\end{figure}

	We plot the cumulative distributions of the DL network sum-rate of our proposed approach in \figref{cdf}.
	To reduce the simulation complexity, we set $M_u = 16 \times 4,\forall u$, $N_{k,u} = 8,\forall \left(k,u\right)\in\mathcal{K}$. The transmit power budget is set to be $40$ dBm.
	Denote the $5\%$ DL sum-rate metric as $R_{0.05}$, i.e., $\mathrm{Pr}\lbrace R \ge R_{0.05}\rbrace \ge 95\%$. As can be seen from \figref{cdf}, the proposed network transmission approach can provide about $116\%$ and $46\%$ performance gains in terms of $R_{0.05}$ over the traditional single-cell and the coordinated baselines, respectively. The performance gains can be attributed to the following two aspects: 1) when the propagation link between a UT and the BS in its cell is blocked, the BSs of other cells in the network can transmit signals to this UT, which can mitigate the blockage effect, resulting in transmission performance gains; 2) when the transmission link state of a UT and the BS in its cell is good, the BSs in other cells in the network can still transmit signals to this UT. The array gain can further lead to performance gains over the traditional approaches.

	\begin{figure*}[!t]
		\centering
		\subfloat[]{\centering\includegraphics[width=0.48\textwidth]{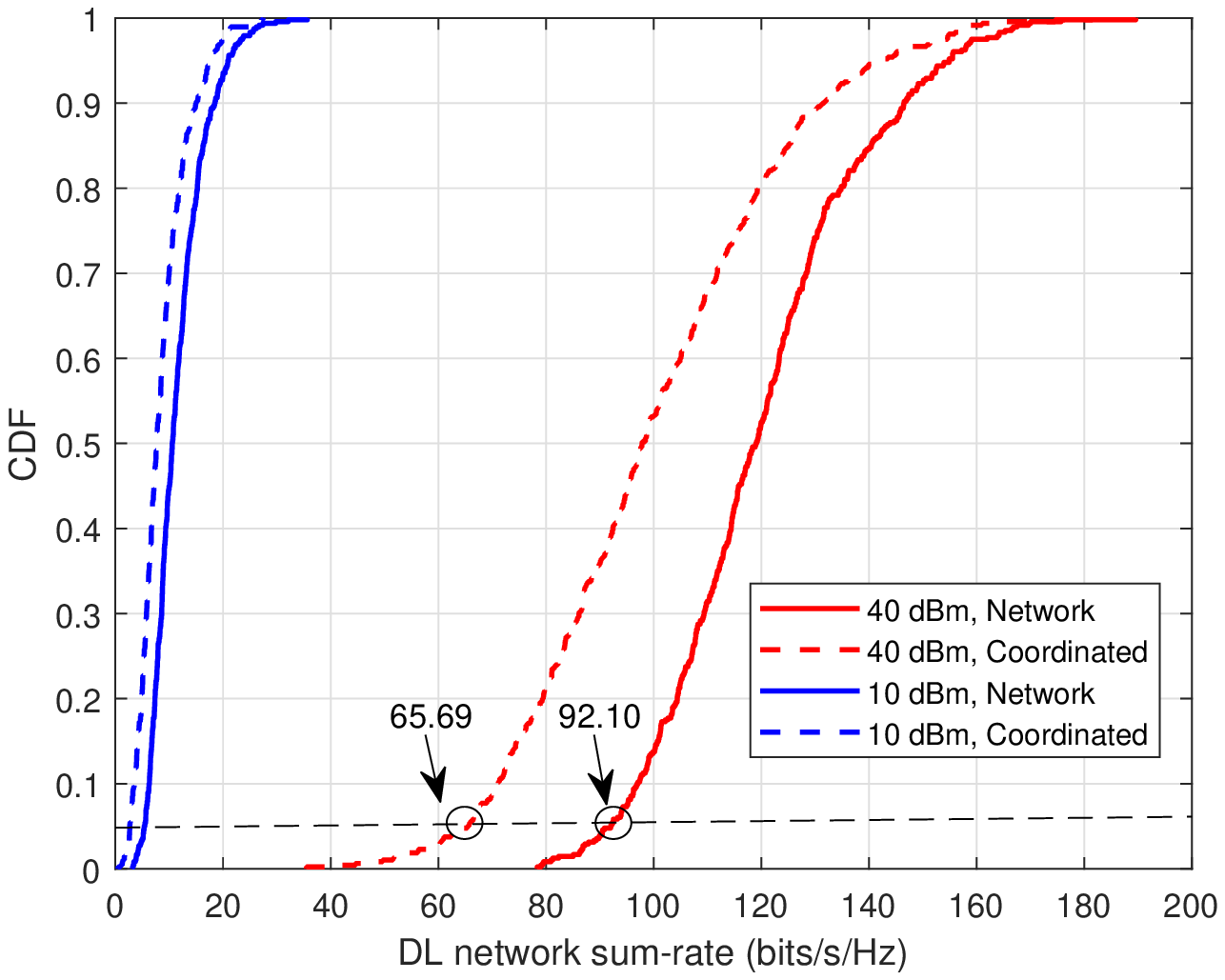}
			\label{cdf_P01}}
		\hfill
		\subfloat[]{\centering\includegraphics[width=0.48\textwidth]{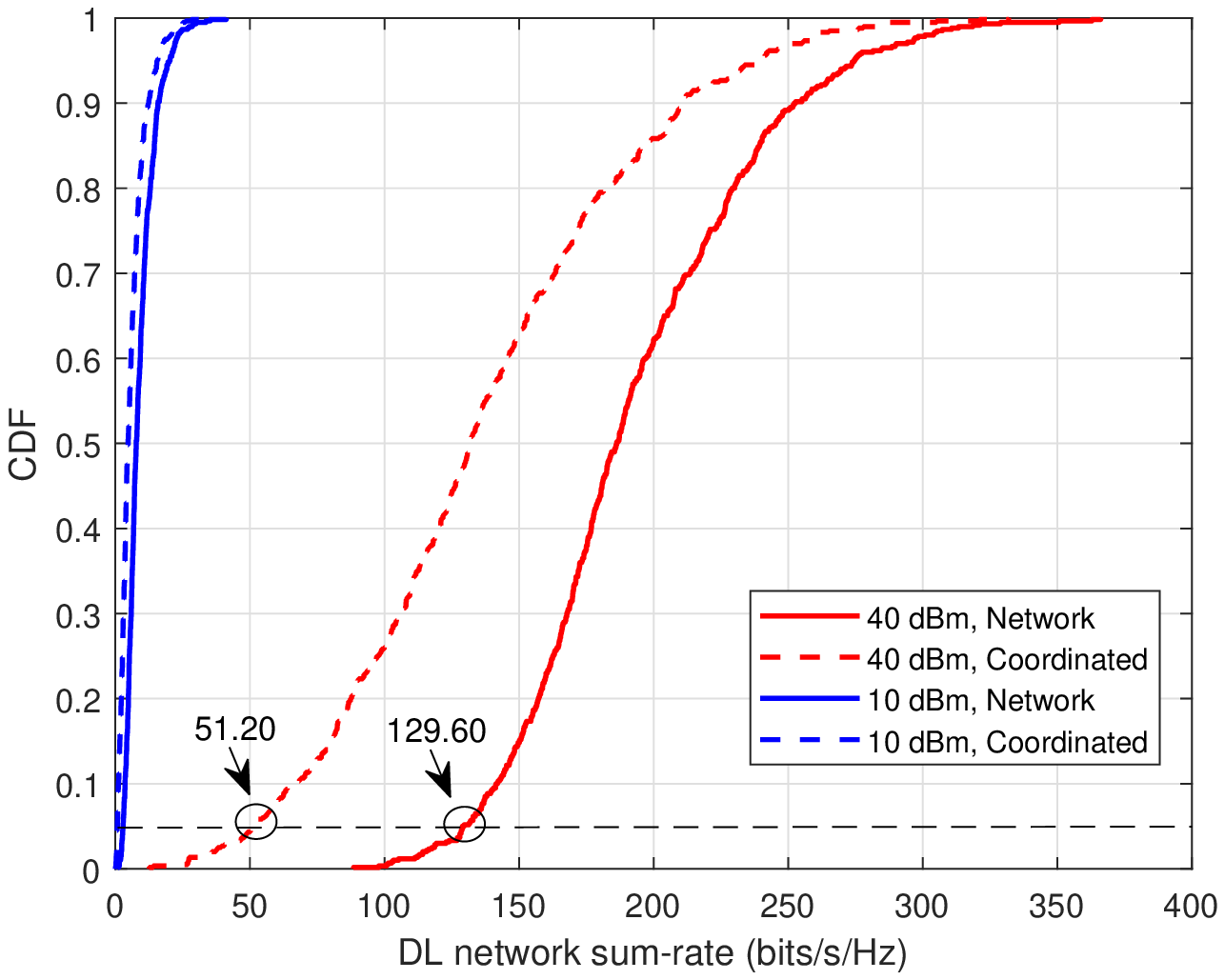}
			\label{cdf_P05}}
		\caption{CDF performance of the DL network sum-rate for the network and coordinated transmission approaches. (a) $28$ GHz, $M_{u} = 16\times 4,\forall u,N_{k,u} =8,\forall \left(k,u\right)\in\mathcal{K},p_{\mathrm{out}} = 0.1$; (b) $100$ GHz, $M_{u} = 16\times 4,\forall u,N_{k,u} =32,\forall \left(k,u\right)\in\mathcal{K},p_{\mathrm{out}} = 0.5$.  }
		\label{cdf_2}
	\end{figure*}

	\figref{cdf_2} further compares the proposed network transmission approach with the coordinated baseline. Firstly, in order to evaluate the effect of blockage on the two approaches, we fix $p_\mathrm{out}$ of formula \eqref{P_out} as $p_\mathrm{out}=0.1$ in Fig. (4a) and $p_\mathrm{out}=0.5$ in Fig. (4b), respectively. For the case with $P_\mathrm{max}= 40$ dBm, it can be observed that the performance gain of the proposed network transmission is more significant in the severer THz blockage scenarios. Notably,
	we can observe the performance gains of $R_{0.05}$ in Fig. (4a) and in Fig. (4b) are $40.2\%$ and $153.1\%$, respectively.
	In addition, the proposed network transmission approach has an average transmission rate gain of $20.1\%$ in Fig. (4a) and $41.2\%$ in Fig. (4b), respectively. With the above observations, we find that the proposed network transmission approach presents significant transmission performance gain over the traditional coordinated one, especially when the channel is severely blocked. 	
	Furthermore, we can also observe that in the case of high transmit power budget $P_{\mathrm{max}}$, the performance gain of the proposed approach is more significant.
	

	\begin{figure*}[!t]
		\centering
		\subfloat[]{\centering\includegraphics[width=0.48\textwidth]{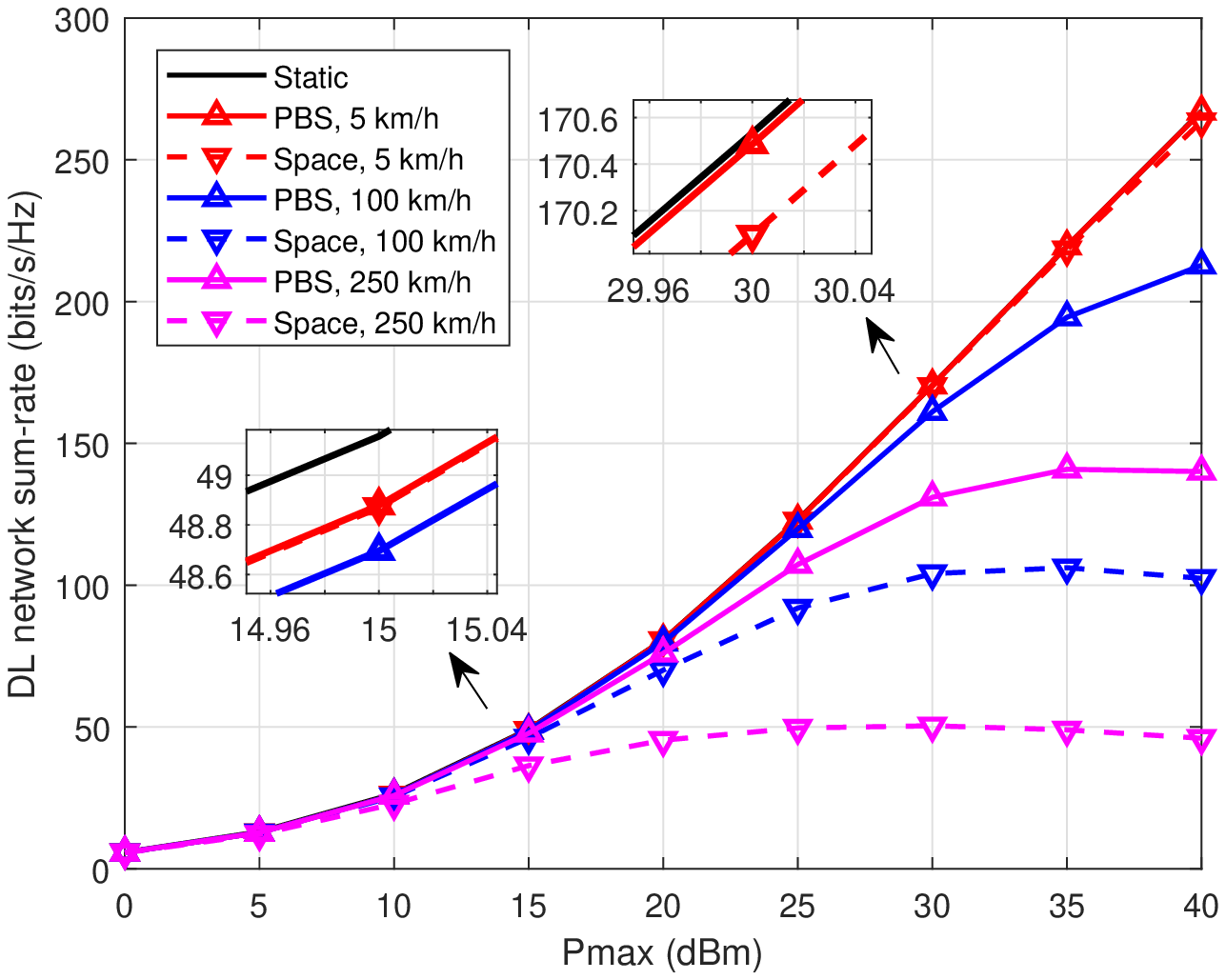}
			\label{PBS_or_Not-a}}
		\hfill
		\subfloat[]{\centering\includegraphics[width=0.48\textwidth]{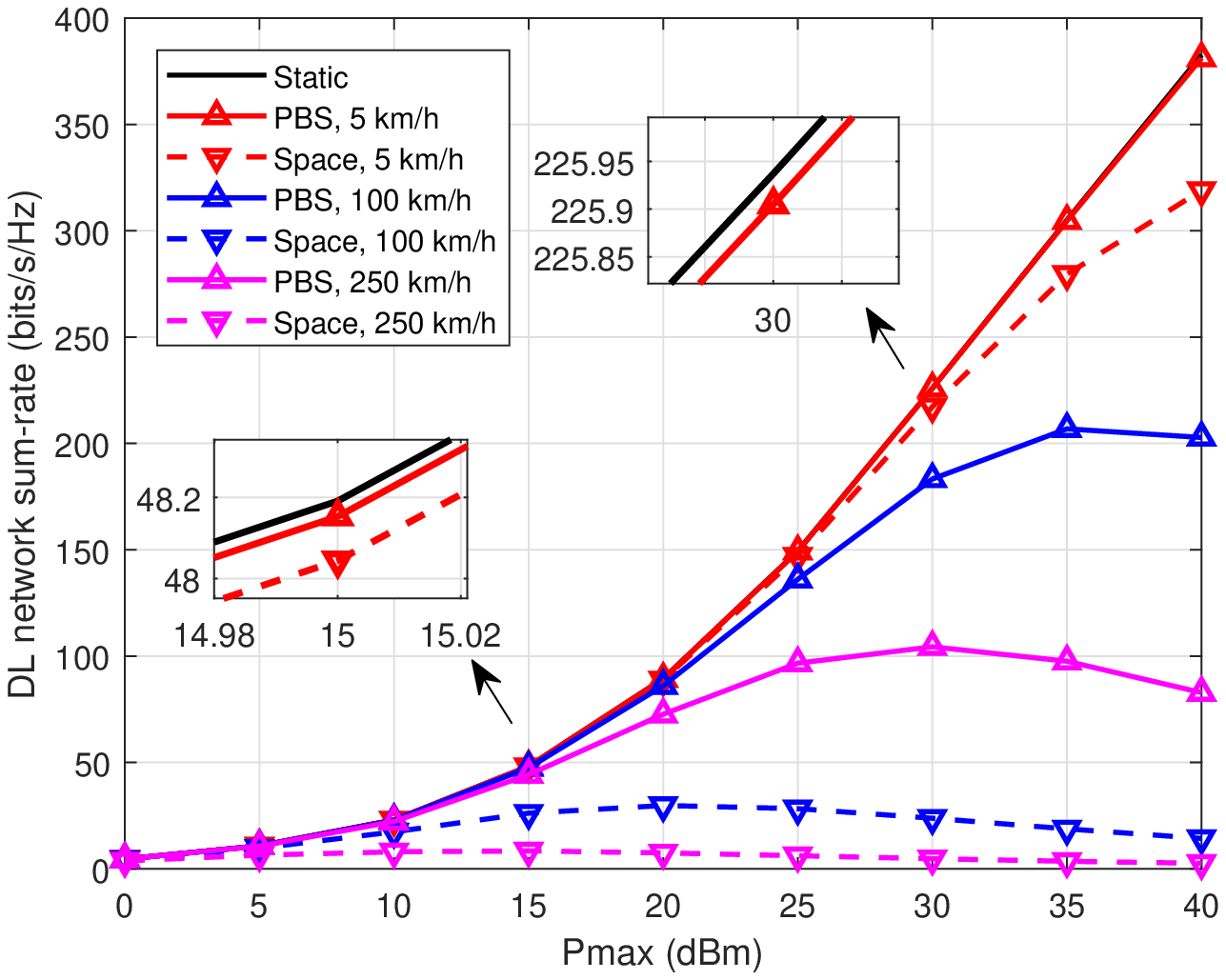}
			\label{PBS_or_Not-b}}
		\caption{Comparison of the DL network sum-rate performance with PBS and the conventional space domain synchronization. (a) $28$ GHz, $N_{k,u} =16,\forall \left(k,u\right)\in\mathcal{K}$; (b) $100$ GHz, $N_{k,u} =64,\forall \left(k,u\right)\in\mathcal{K}$.  }
		\label{PBS_or_Not}
	\end{figure*}

	\figref{PBS_or_Not} evaluates the performance of our proposed approach under different mobility scenarios. We adopt a tight universal upper bound of the ICI power due to the Doppler effects as in \cite{OFDM01Li,OFDM02Stamoulis}. We compare the DL network sum-rate performance with PBS and the conventional space domain synchronization \cite{Morelli07Synchronization}. The ideal case where channels are static and can be perfectly obtained is presented as the comparison benchmark. We can observe that the PBS scheme outperforms the conventional one, especially in high mobility scenarios, which demonstrates the effectiveness of our proposed transmission approach for mobility enhancement over the mmWave and THz bands.

	\section{Conclusion}\label{sec:conclusion}
	We have investigated network transmission with only the statistical CSI at the BSs to address the challenging issues of mobility enhancement and blockage mitigation in massive MIMO mmWave/THz communications.
    Considering the mmWave/THz propagation characteristics, we have proposed to apply the PBS scheme for network massive MIMO to mitigate the channel Doppler and delay dispersions, and established a transmission model accordingly.
    Our design goal has been successfully transformed into the maximization of the DL network transmission sum-rate.
    We have shown the optimality of the beam domain transmission, and demonstrated
    that different BSs can work individually when sending signals to UTs. We have also developed a
    low-complexity iterative algorithm with guaranteed convergence based on the CCCP and random matrix theory.
	The KKT conditions have been used to solve the problem in each iteration. We have shown that the solution of the KKT conditions has a classical water-filling structure. As a result, an iterative water-filling power allocation algorithm with guaranteed convergence has been proposed.
	Numerical results have demonstrated the effectiveness of the proposed transmission approach for alleviating the blockage effects and providing mobility enhancement over the mmWave and THz bands.

	\appendices
	\section{Proof of \thref{prop_optimal_dl}}\label{optimal_dl}
	
	From \eqref{K^dl} and \eqref{Xi_dl}, we can observe that $\bK_{k,u}$ is a diagonal matrix, and the off-diagonal entries of $\bQ_{i,j},\forall\left(i,j\right)\neq \left(k,u\right)$, will not affect the value of $\bK_{k,u}$. Using a method similar to that in \cite{You2020energy},
    we define a diagonal matrix $\boldsymbol{\Upsilon}_m\in\bbR^{M_{\mathrm{tot}}\times M_{\mathrm{tot}}}$ whose diagonal entries are $1$ except the $\left(m,m\right)$-th entry, which is $-1$.
    The off-diagonals in the $m$-th row and $m$-th column of matrices $\boldsymbol{\Upsilon}_m\bQ_{k,u}\boldsymbol{\Upsilon}_m$ and $\bQ_{k,u}$ have the same absolute values but opposite signs, and the other entries of matrix $\boldsymbol{\Upsilon}_m\bQ_{k,u}\boldsymbol{\Upsilon}_m$ are equal to those of $\bQ_{k,u}$.

    Note that $\boldsymbol{\Upsilon}_m\bQ_{k,u}\boldsymbol{\Upsilon}_m\odot\bI=\bQ_{k,u}\odot\bI$, we then have		
	\begin{align}
	R\left(\bQ_{1,1},\cdots,\bQ_{K_U,U}\right)=&\sum_{u=1}^{U}\sum_{k=1}^{K_u}\mathbb{E}\left\lbrace\logdet{\bI+\bK_{k,u}^{-1}\bG_{k,u}\bQ_{k,u}\bG_{k,u}^H}\right\rbrace\ntb
	\equaa&\sum_{u=1}^{U}\sum_{k=1}^{K_u}\mathbb{E}\left\lbrace\logdet{\bI+\bK_{k,u}^{-1}\bG_{k,u}\boldsymbol{\Upsilon}_m\bQ_{k,u}\boldsymbol{\Upsilon}_m\bG_{k,u}^H}\right\rbrace\ntb
	=&R\left(\boldsymbol{\Upsilon}_m\bQ_{1,1}\boldsymbol{\Upsilon}_m,\cdots,\boldsymbol{\Upsilon}_m\bQ_{K_U,U}\boldsymbol{\Upsilon}_m\right).
	\end{align}
Because of the independence of the columns of $\bG_{k,u}$  and their symmetric distributions, reserving the sign of the $m$-th column does not change its distribution. Consequently, $\left(\mathrm{a}\right)$ is established.

Note that compared to $\bQ_{k,u}$, the effect of computing $\frac{1}{2}\left(\bQ_{k,u}+\boldsymbol{\Upsilon}_m\bQ_{k,u}\boldsymbol{\Upsilon}_m\right)$ is to null the off-diagonal entries in the $m$-th row and $m$-th column of $\bQ_{k,u}$.
Moreover, because of the concavity of $\mathrm{log}\left(\det\left(\cdot\right)\right)$, using Jensen's inequality, we have
	\begin{align}\label{Jensen}
	&R\left(\frac{1}{2}\left(\bQ_{1,1}+\boldsymbol{\Upsilon}_m\bQ_{1,1}\boldsymbol{\Upsilon}_m\right),\cdots,\frac{1}{2}\left(\bQ_{K_U,U}+\boldsymbol{\Upsilon}_m\bQ_{K_U,U}\boldsymbol{\Upsilon}_m\right)\right)\ntb
	=&\sum_{u=1}^{U}\sum_{k=1}^{K_u}\mathbb{E}\left\lbrace\logdet{\bI+\frac{1}{2}\bK_{k,u}^{-1}\bG_{k,u}\left(\bQ_{k,u}+\boldsymbol{\Upsilon}_m\bQ_{k,u}\boldsymbol{\Upsilon}_m\right)\bG_{k,u}^H}\right\rbrace\ntb
	\ge&\frac{1}{2}\sum_{u=1}^{U}\sum_{k=1}^{K_u}\mathbb{E}\left\lbrace\logdet{\bI+\bK_{k,u}^{-1}\bG_{k,u}\bQ_{k,u}\bG_{k,u}^H}\right\rbrace\ntb &\quad+\frac{1}{2}\sum_{u=1}^{U}\sum_{k=1}^{K_u}\mathbb{E}\left\lbrace\logdet{\bI+\bK_{k,u}^{-1}\bG_{k,u}\boldsymbol{\Upsilon}_m\bQ_{k,u}\boldsymbol{\Upsilon}_m\bG_{k,u}^H}\right\rbrace\ntb
	=&\frac{1}{2}R\left(\bQ_{1,1},\cdots,\bQ_{K_U,U}\right)+ \frac{1}{2}R\left(\boldsymbol{\Upsilon}_m\bQ_{1,1}\boldsymbol{\Upsilon}_m,\cdots,\boldsymbol{\Upsilon}_m,\bQ_{K_U,U}\boldsymbol{\Upsilon}_m\right)\ntb
	=&R\left(\bQ_{1,1},\cdots,\bQ_{K_U,U}\right).
	\end{align}
Thus, we can increase the DL network transmission sum-rate $R$ by nulling the off-diagonal entries in the $m$-th row and $m$-th column of $\bQ_{1,1},\cdots,\bQ_{K_U,U}$. Repeating this process for $m$ from $1$ to $M_{\mathrm{tot}}$, we draw the conclusion that $R$ is maximized when $\bQ_{1,1},\cdots,\bQ_{K_U,U}$ are all diagonal. This concludes the proof.

	\section{Proof of \thref{prop_Lambda_dl}}\label{Lambda_dl}
	We define the Lagrangian function of the problem in \eqref{OP4_dl} as follows
	\begin{align}
	\digamma &= \sum_{u=1}^{U}\sum_{k=1}^{K_u}\left\lbrace \overline{f}_{k,u}^+\left(\bLambda\right)-f_{k,u}^-\left(\bLambda^{\left(\ell\right)}\right)-\mathrm{tr}\left(\boldsymbol{\Delta}_{k,u}^{\left(\ell\right)}\left(\bLambda_{k,u}-\bLambda_{k,u}^{\left(\ell\right)}\right)\right)\right\rbrace\ntb
	&\quad +\sum_{u=1}^{U}\sum_{k=1}^{K_u}\mathrm{tr}\left(\mathbf{\Psi}_{k,u}\bLambda_{k,u}\right)- \sum_{v=1}^{U}\mu_v\left(\mathrm{tr}\sum_{u=1}^{U}\sum_{k=1}^{U_k}\left(\bE_v\bLambda_{k,u}\right)-P_v\right),
	\end{align}
	where the Lagrange multipliers $\mathbf{\Psi}_{k,u}\succeq \mathbf{0}$ and $\mu_v\ge 0$ are related to the problem constraints.
	
	The gradient of $\overline{f}_{k,u}^+\left(\bLambda\right)$ with respect to $\bLambda_{k,u}$ can be derived from \eqref{f_k_u_de^+^dl} as
	\begin{align}
	\frac{\partial}{\partial \bLambda_{k,u}}\overline{f}_{k,u}^+\left(\bLambda\right) &= \left(\bI_{M_\mathrm{tot}} + \mathbf{\Gamma}_{k,u}\bLambda_{k,u}\right)^{-1}\mathbf{\Gamma}_{k,u} \ntb
	&\quad+\sum_{m,n}\frac{\partial \overline{f}_{k,u}^+\left(\bLambda\right)}{\partial\left[\mathbf{\Xi}_{k,u}\left(\bPhi_{k_u}^{-1}\bLambda_{k,u}\right)\right]_{m,n}}\frac{\partial\left[\mathbf{\Xi}_{k,u}\left(\bPhi_{k,u}^{-1}\bLambda_{k,u}\right)\right]_{m,n}}{\partial \bLambda_{k,u}}\ntb
	&\quad+\sum_{m,n}\frac{\partial \overline{f}_{k,u}^+\left(\bLambda\right)}{\partial\left[\bPi_{k,u}\left(\widetilde{\bPhi}_{k,u}^{-1}\bK_{k,u}^{-1}\right)\right]_{m,n}}\frac{\partial\left[\bPi_{k,u}\left(\widetilde{\bPhi}_{k,u}^{-1}\bK_{k,u}^{-1}\right)\right]_{m,n}}{\partial\bLambda_{k,u}}.
	\end{align}
    Following an approach similar to that for proving Theorem 4 in \cite{Lu16Free}, we have
	\begin{align}
	\frac{\partial \overline{f}_{k,u}^+\left(\bLambda\right)}{\partial\left[\mathbf{\Xi}_{k,u}\left(\bPhi_{k,u}^{-1}\bLambda_{k,u}\right)\right]_{m,n}}=0,\\
	\frac{\partial \overline{f}_{k,u}^+\left(\bLambda\right)}{\partial\left[\bPi_{k,u}\left(\widetilde{\bPhi}_{k,u}^{-1}\left(\bK_{k,u}\right)^{-1}\right)\right]_{m,n}}=0,
	\end{align}
	which further leads to
	\begin{align}\label{partial_f_k_u_de_dl}
	\frac{\partial}{\partial \bLambda_{k,u}}\overline{f}_{k,u}^+\left(\bLambda\right)=\left(\bI_{M_\mathrm{tot}}+\mathbf{\Gamma}_{k,u}\bLambda_{k,u}\right)^{-1}\mathbf{\Gamma}_{k,u}.
	\end{align}
	In addition, we can obtain the gradient of $\overline{f}_{k,u}^+\left(\bLambda\right)$ over $\bLambda_{i,j}$, $\forall \left(i,j\right)\neq \left(k,u\right)$, as
	\begin{align}\label{partial_f_i_j_de_dl}
	\frac{\partial}{\partial \bLambda_{i,j}}\overline{f}_{k,u}^+\left(\bLambda\right) = \sum_{n=1}^{N_{k,u}}\frac{\hatbR_{i,j,n}}{\widetilde{\gamma}_{i,j,n}+\sigma^2+\mathrm{tr}\left(\hatbR_{i,j,n}\bLambda_{\backslash \left(i,j\right)}\right)}.
	\end{align}
	Then, from \eqref{partial_f_k_u_de_dl} and \eqref{partial_f_i_j_de_dl}, we have
	\begin{align}
	\frac{\partial}{\partial \bLambda_{a,b}}\sum_{u=1}^{U}\sum_{k=1}^{K_u}\overline{f}_{k,u}^+\left(\bLambda\right) &= \left(\bI_{M_\mathrm{tot}} + \mathbf{\Gamma}_{a,b}\bLambda_{a,b}\right)^{-1}\mathbf{\Gamma}_{a,b} \ntb &\quad+\sum_{\left(k,u\right)\neq \left(a,b\right)}\sum_{n=1}^{N_{k,u}}\frac{\hatbR_{k,u,n}}{\widetilde{\gamma}_{k,u,n}+\sigma^2+\mathrm{tr}\left(\hatbR_{k,u,n}\bLambda_{\backslash{\left(k,u\right)}}\right)}.
	\end{align}
    Owing to the concavity of $\overline{f}_{k,u}^+\left(\bLambda\right)$ over $\bLambda$, the KKT conditions of \eqref{OP4_dl} are
	\begin{subequations}\label{KKT}
		\begin{align}
		\frac{\partial \digamma}{\partial \bLambda_{k,u}^{\left(\ell+1\right)}}=\mathbf{0}, & \label{first_KKT_dl}\\
		\mathrm{tr}\left(\bPsi_{k,u}^{\left(\ell+1\right)}\bLambda_{k,u}^{\left(\ell+1\right)}\right)=0,&\quad\bPsi_{k,u}^{\left(\ell+1\right)}\succeq \mathbf{0},\quad\bLambda_{k,u}^{\left(\ell+1\right)}\succeq \mathbf{0},\label{second_KKT_dl}\\
		\mu_v\left(\tr{\sum_{u=1}^{U}\sum_{k=1}^{K_u}\left(\bE_v\bLambda_{k,u}\right)-P_v}\right)=0,&\quad\mu_v\ge 0,\quad\forall\left(k,u\right)\in\mathcal{K}.\label{thrid_KKT_dl}
		\end{align}
	\end{subequations}
    Utilizing the concavity of problem \eqref{OP4_dl}, we can get the optimal solution $\bLambda_{k,u}^{\left(\ell+1\right)}$ by solving the corresponding KKT conditions.	
    We reformulate the first KKT condition in \eqref{first_KKT_dl} as
	\begin{align}\label{first_KKT_dl_2}
	\frac{\partial \digamma}{\partial \bLambda_{k,u}^{\left(\ell+1\right)}} &= \left(\bI_{M_\mathrm{tot}} + \mathbf{\Gamma}_{k,u}^{\left(\ell+1\right)}\bLambda_{k,u}^{\left(\ell+1\right)}\right)^{-1}\mathbf{\Gamma}_{k,u}^{\left(\ell+1\right)}  - \boldsymbol{\Delta}_{k,u}^{\left(\ell\right)} +\bPsi_{k,u}^{\left(\ell+1\right)} - \mu_v^{\left(\ell\right)}\bE_v \ntb
	&\quad 	+\sum_{\left(i,j\right)\neq \left(k,u\right)}\sum_{n=1}^{N_{i,j}}\frac{\hatbR_{i,j,n}}{\widetilde{\gamma}^{\left(\ell+1\right)}_{i,j,n}+1+  \mathrm{tr}\left(\hatbR_{i,j,n}\bLambda_{\backslash{\left(i,j\right)}}^{\left(\ell +1\right)}\right)}  = \mathbf{0}.
	\end{align}
	It can be found that the KKT conditions in \eqref{KKT} equal to those of the following problem
	\begin{align}\label{OP5_dl_2}
	\argmax{\bLambda}\;&\sum_{u=1}^{U}\sum_{k=1}^{K_u}
\Big\{ \logdet{\bI_{M_\mathrm{tot}}+\mathbf{\Gamma}_{k,u}\bLambda_{k,u}}
+\logdet{\widetilde{\mathbf{\Gamma}}_{k,u}+\bK_{k,u}\left(\bLambda\right)}  \ntb
&\qquad\qquad-\mathrm{tr}\left(\boldsymbol{\Delta}_{k,u}^{\left(\ell\right)}\bLambda_{k,u}\right)\Big\}\ntb
	\mathrm{s.t.}\quad&\tr{\sum_{u=1}^{U}\sum_{k=1}^{K_u}\left(\bE_v \bLambda_{k_u}\right)}\le P_v,\quad\forall v\in\mathcal{U},\ntb
	& \bLambda_{k,u} \succeq \mathbf{0},\; \bLambda_{k,u}\; \mathrm{diagonal},\qquad\forall \left(k,u\right)\in\mathcal{K}.
	\end{align}
    Solving the corresponding KKT conditions, we have
	\begin{align}\label{lambda_dl_2}
		\left\{
			{\begin{array}{*{20}{c}}
			{
			\frac{\gamma_{k,u,m}^{\left(\ell\right)}}{1+\gamma_{k,u,m}^{\left(\ell\right)}\lambda_{k,u,m}^{\left(\ell+1\right)}} +\sum\limits_{\left(i,j\right)\neq \left(k,u\right)}\sum\limits_{n=1}^{N_{i,j}}\frac{\hat{r}_{i,j,m,n}}{\widetilde{\gamma}_{i,j,n}^{\left(\ell\right)}+\sigma^2+\mathrm{tr}\left(\hatbR_{i,j,n}\bLambda_{\backslash \left(i,j\right)}^{\left(\ell+1\right)}\right)}=\delta_{k,u,m}^{\left(\ell\right)}+\mu_v,
			\mu_v<\chi_{k,u,m}^{\left(\ell+1\right)}-\delta_{k,u,m}^{\left(\ell\right)},
			}\\
			{
			\lambda_{k,u,m}^{\left(\ell+1\right)}=0,\qquad\qquad\qquad\qquad\qquad\qquad\qquad\qquad\qquad\qquad\qquad
			\mu_v\ge\chi_{k,u,m}^{\left(\ell+1\right)}-\delta_{k,u,m}^{\left(\ell\right)}
		    },
			\end{array}}
			\right.
	\end{align}
	where the auxiliary variable $\chi_{k,u,m}^{\left(\ell+1\right)}$ is expressed as
	\begin{align}\label{chi_dl_2}
	\chi_{k,u,m}^{\left(\ell+1\right)}=	\gamma_{k,u,m}^{\left(\ell+1\right)}+ \sum\limits_{\left(i,j\right)\neq \left(k,u\right)}\sum\limits_{n=1}^{N_{i,j}}\frac{\hat{r}_{i,j,m,n}}{\widetilde{\gamma}_{i,j,n}^{\left(\ell+1\right)} +\sigma^2+ \sum\limits_{\left(p,q,m'\right) \in \mathcal{S}_{k,u,m,i,j}} \hat{r}_{i,j,m',n}\lambda_{p,q,m'}^{\left(\ell+1\right)}}.
	\end{align}
	This concludes the proof.



\end{document}